\documentclass[aps,reprint,prb,longbibliography,superscriptaddress]{revtex4-2}

\usepackage{amsmath}
\usepackage{amssymb}
\usepackage{mathtools}
\usepackage{amsthm}
\usepackage{bbm}
\usepackage{amsfonts}
\usepackage{braket}
\usepackage{algorithmic}
\usepackage{lipsum}
\usepackage{graphicx}
\usepackage{xcolor}
\usepackage{xfrac}
\usepackage{calc}
\usepackage{accents}
\newcommand{\dbtilde}[1]{\accentset{\approx}{#1}}

\usepackage{amsthm,letltxmacro,changepage}

\DeclareMathOperator{\tr}{tr}
\DeclareMathOperator{\TS}{\hat{\mathbf{S}}}
\DeclareMathOperator{\Doa}{\Downarrow}
\DeclareMathOperator{\Upa}{\Uparrow}
\DeclareMathOperator{\doa}{\downarrow}
\DeclareMathOperator{\upa}{\uparrow}
\DeclareMathOperator{\ud}{\uparrow\downarrow}

\newcommand{\cfig}[1]{Fig.~\ref{#1}}
\newcommand{\ceqn}[1]{Eq.~\ref{#1}}
\newcommand{\capp}[1]{Appendix~\ref{#1}}
\newcommand{\cref}[1]{Ref.~\cite{#1}}
\newcommand{\csec}[1]{Sec.~\ref{#1}}

\newcommand{\cact}[1]{\hyperref[#1]{#1}}
\newcommand{\ignore}[1]{}

\DeclareUnicodeCharacter{0301}{*************************************}

\DeclareMathOperator{\eff}{\textrm{eff}}

\DeclareMathOperator{\dd}{\text{d}\!}

\DeclareMathOperator{\hc}{{\rm H.c.}}

\DeclareMathOperator{\normL}{\vert\!\vert} 
\DeclareMathOperator{\normR}{\vert\!\vert}
\newcommand{\norm}[1]{\normL #1 \normR}
\newcommand{\vek}[1]{\mathbf{#1}} % vector style

\def\getfirst#1.#2\relax{#1}
\newcommand{\rev}[1]{\textcolor{black}{#1}}

\newcommand{\aim}{AIM} % single anderson impurity model
\newcommand{\gml}{UML} % generative model learning for molecular electronics
\newcommand{\fp}{FP} % fixed point
\newcommand{\qn}{QN} % quantum numbeer
\definecolor{commcol}{RGB}{0,0,0}
\definecolor{new_text}{RGB}{0,0,0}

\begin{document}

%########################
%########################

\title{Unsupervised Learning of Effective Quantum Impurity Models}

\author{Jonas B. Rigo}
\email[]{j.rigo@fz-juelich.de}
\affiliation{Forschungszentrum J\"{u}lich GmbH, Peter Gr\"{u}nberg Institute,
	Quantum Control, 52425 J\"{u}lich, Germany}
 	\affiliation{School of Physics, University College Dublin, Belfield, Dublin 4, Ireland}
	\affiliation{Centre for Quantum Engineering, Science, and Technology, University College Dublin, Dublin 4, Ireland}
\author{Andrew K. Mitchell}
\email[]{andrew.mitchell@ucd.ie}
	\affiliation{School of Physics, University College Dublin, Belfield, Dublin 4, Ireland}
	\affiliation{Centre for Quantum Engineering, Science, and Technology, University College Dublin, Dublin 4, Ireland}
%########################
%########################

\begin{abstract}
Generalized quantum impurity models -- which feature a few localized and strongly-correlated degrees of freedom coupled to itinerant conduction electrons -- describe diverse physical systems, from magnetic moments in metals to nanoelectronics quantum devices such as quantum dots or single-molecule transistors. Correlated materials can also be understood as self-consistent impurity models through dynamical mean field theory. Accurate simulation of such models is challenging, especially at low temperatures, due to many-body effects from electronic interactions, resulting in strong renormalization. In particular, the interplay between local impurity complexity and Kondo physics is highly nontrivial. A common approach, which we further develop in this work, is to consider instead a simpler effective impurity model that still captures the low-energy physics of interest. The mapping from bare to effective model is typically done perturbatively, but even this can be difficult for complex systems, and the resulting effective model parameters can anyway be quite inaccurate. Here we develop a non-perturbative, unsupervised machine learning approach to systematically obtain low-energy effective impurity-type models, based on the renormalization group framework. The method is shown to be general and flexible, as well as  accurate and systematically improvable. We benchmark the method against exact results for the Anderson impurity model, and provide an outlook for more complex models beyond the reach of existing methods.
\end{abstract}
\maketitle

%########################
% Introduction
%########################

In the field of condensed matter physics,
the microscopic Hamiltonian describing a quantum many-body system (the \textit{bare model}) is in many cases well known. However, for interacting systems, the complexity of these models grows quickly with the number of quantum degrees of freedom (for example orbitals or spins), such that a brute force solution becomes analytically and/or numerically intractable for many realistic scenarios of interest. The challenge in many-body theory is therefore not in writing down the bare model, but rather in solving it. 
However, in many situations it is not necessary to consider the entire configuration space of the system, because only a (relatively) small \textit{active} subspace controls the phenomena of interest \cite{thouless2014quantum}. Thus, \textit{effective models} can be devised, that faithfully capture the phenomena of interest, while only keeping the active part of the configuration space. Such effective models have a reduced complexity and increased expressiveness. The challenge in many-body theory can therefore be restated as one of finding the best \textit{solvable} model that describes approximately but accurately the physics of interest.

% ab-initio downfolding
As an example, it is often more convenient to analyse an effective lattice model in second quantized form than the \textit{ab initio} treatment involving an all-orbital description of the constituent atoms \cite{hubbard1963electron,kanamori1963electron,calvo2024theoretical}.
Methods such as the constrained random-phase approximation \cite{aryasetiawan2004frequency,csacsiouglu2011effective}, coupled cluster downfolding \cite{bauman2022coupled}, density matrix downfolding \cite{zheng2018real} or the Pariser-Parr-Pople model for molecules \cite{pariser1953semi,*pariser1953semi2} were devised to systematically eliminate inactive degrees of freedom and account for them with renormalized parameters of a reduced model.

%%%%%%%%%%%%%%%%%%%%%%%%%%%%%%%%%%%%%%%%%
\begin{figure}[b]
	\centering
	\includegraphics[width=0.82\columnwidth]{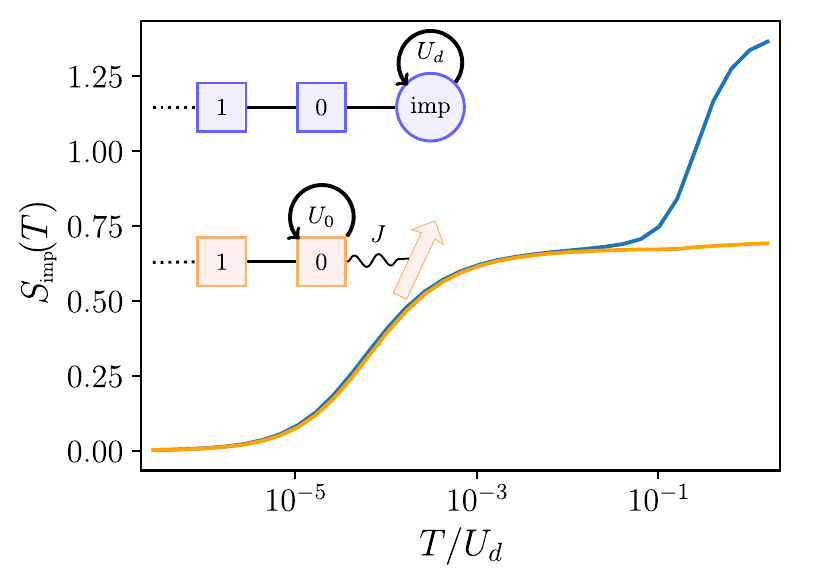}
	\caption{Schematic of the simplest bare (blue) and effective (orange) impurity models, together with an illustrative comparison of their impurity entropies after model optimization.
 }	\label{fig:schematic}
\end{figure}
%%%%%%%%%%%%%%%%%%%%%%%%%%%%%%%%%%%%%%%%%

% Machine leraning inspiration
These methods have in common that they require some prior knowledge about the bare model to derive the effective model. This situation shares some similarity with that of the \textit{inverse problem} \cite{nguyen2017inverse} encountered in statistical inference or machine learning \cite{bengio}, that seeks to infer a probabilistic model from observed data. 
% Application examples
Recently machine learning techniques have been explored to construct effective Hamiltonians from simulated data \cite{bhat2020machine,wang2021machine,sawaya2022constructing,nakhaee2020machine} or from experimentally measured data 
\cite{prufer2020experimental,yu2021learning,schuster2022learning}.
% NISQ
Another field where the inference of Hamiltonians has gained considerable relevance, is in the analogue simulation of quantum Hamiltonians on quantum hardware \cite{buluta2009quantum,*georgescu2014quantum,barthelemy2013quantum,*hensgens2017quantum,pouse2023quantum}. Indeed, machine learning inspired process- and Hamiltonian-tomography methods have been developed to infer the precise Hamiltonian that NISQ devices actually simulate, given hardware imperfections \cite{anshu2021sample,boulant2003robust,zhang2014quantum,kieferova2017tomography,olsacher2022digital,schmitt2023reconstructing}.

A good effective Hamiltonian must not just reproduce desired properties of the bare model, it must also be meaningfully simpler than the bare model. The canonical method to systematically eliminate degrees of freedom and obtain  effective models is the renormalization group (RG) \cite{wilson1971renormalization,glazek1993renormalization,kopietz2010introduction}. On the level of the thermal density matrix a single RG transformation acts as the partial trace over the high energy degrees of freedom \cite{beny2015information}. In the context of quantum physics, a central concept from information theory is that thermal states encode the corresponding Hamiltonian \cite{garrison2018does}, such that comparing the ensemble of thermal states from two systems is equivalent to comparing the two Hamiltonians. This provides a way to measure the `difference' between a Hamiltonian before and after renormalization. By minimizing the distance in Hamiltonian space, one can therefore in principle optimize a simplified effective model to best approximate some more complicated bare Hamiltonian after renormalization. We rigorously derive such an approach in this paper for a specific class of systems known as `quantum impurity' models \cite{hewson1993}. These models involve localized interacting quantum degrees of freedom (the `impurity' $\hat{H}^{\rm imp}$) coupled (throguh $\hat{H}^{\rm hyb}$) to one or more continuum baths of noninteracting conduction electrons ($\hat{H}^{\rm bath}$), as described generically by:
\begin{equation}\label{eq:Hqim}
\hat{H}=\hat{H}^{\rm imp} + \hat{H}^{\rm bath} + \hat{H}^{\rm hyb} \;.
\end{equation}
Generalized quantum impurity models describe semiconductor quantum dot devices \cite{goldhaber1998kondo} and complex single-molecule junctions \cite{Mitchell2017}. They also underpin our understanding of correlated materials through dynamical mean field theory \cite{georges1996dynamical}. Developing a strategy for systematically and accurately deriving simple effective models to better understand and simulate complex systems is the ultimate application of this work.

A prerequisite for the design and use of complex quantum nanoelectronics devices with advanced functionality beyond the classical paradigm (for example in quantum metrology \cite{mihailescu2024quantum,mihailescu2024multiparameter,*mihailescu2023thermometry}), is a fundamental understanding of their low-temperature correlated electron physics and quantum transport properties. However, this is a notoriously difficult theoretical challenge because of the subtle interplay between the orbital and spin complexity of the nanostructure, determined by its structure and chemistry; strong electron interactions due to quantum confinement; and the coupling to $\sim 10^{23}$ conduction electrons in the external circuit. This results in nontrivial quantum phenomena such as Coulomb blockade \cite{Park2002}, various forms of Kondo effect \cite{Park2002,Scott2010,Mitchell2017}, and quantum interference \cite{guedon2012observation,sen2023many} -- all of which strongly affect low-temperature electronic conductance through the device, and hence its functionality. As with coupled quantum dot devices \cite{keller2014emergent,iftikhar2018tunable,*mitchell2016universality,SO5kondo}, entangled spin and charge degrees of freedom can give rise to new physics in single molecule junctions. It is therefore a formidable task to derive simplified effective models that can still describe this range of physics in these kinds of quantum device.

A basic illustrative example is given in Fig.~\ref{fig:schematic}. Here we show the temperature-dependence of a physical observable, in this case the entropy $S(T)$, for a bare quantum impurity model (blue), compared with that of a simpler effective model (orange). One can immediately identify that the effective model has lower complexity than the bare one, because in the high-temperature limit it has a lower entropy -- signifying that the effective model has a smaller number of active degrees of freedom than the bare model. At high temperatures, physical properties of the two models are distinctly different. But below some low-energy scale, the behavior of the two models becomes essentially indistinguishable. The low-energy physics is perfectly captured by the simpler effective model. We will return to the specific details of this example later.

In the context of molecular electronics, the complex microscopic models describing single-molecule junctions can be mapped to much simpler effective two-channel Kondo models using a perturbative approach, as described in Refs.~\cite{Mitchell2017,sen2023many}. This method captures simultaneously the effect of quantum interference and Kondo physics -- but is inevitably approximate due to the perturbative treatment. Since thermodynamic observables flow under RG and the bare and effective model are defined at different energy scales, it is also not \textit{a priori} clear whether such minimal Kondo models are sufficiently general to reproduce local observables of interest in the bare model \cite{rigo2020machine}. 
In this paper we show how an RG analysis of effective interaction terms can be used to determine thermodynamic observables that are comparable across different energy scales. This allows us to introduce a novel machine learning (ML) methodology to derive accurate effective models for complex quantum impurity problems, that works by optimizing generalized (minimally constrained) models, and ensuring that local observables are correctly reproduced. We show by way of explicit examples that the low-energy Kondo physics is simultaneously captured.
The parameters of the effective model are optimized by minimizing the Kullback-Leibler divergence (KLD) \cite{Amari2016} that compares its ensemble of thermal states to that of the bare model. Information on the target is extracted from a numerical simulation of the system of interest. However, the full solution of the bare model is \textit{not} required: minimization of the KLD requires only an estimation of thermal expectation values corresponding to specific local operators at relatively high temperatures. This can be achieved with any suitable quantum impurity solver \cite{SETH2016274,KWW,bulla2008numerical,mitchell2014generalized,*stadler2016interleaved}. Furthermore, the KLD can be shown to be \textcolor{new_text}{convex under reasonable assumptions (see \capp{app:convexity})} and its gradient is known analytically in closed form \cite{grad}, which makes it an ideal optimization problem. \textcolor{new_text}{We refer to this method as \textit{`unsupervised model learning'} (\gml{})}. We demonstrate the efficacy of the \gml{} method by application to the Anderson impurity model (\aim{}) \cite{hewson1993} and obtain some new non-perturbative results for this old problem. Finally we give an outlook to the application of this framework to more complex problems, where the development of tractable effective models is essential for the study of nontrivial correlated electron physics at low temperatures.

%########################
% Method
%########################

\section{Overview of the Method}\label{sec:method}
Unsupervised learning is a type of ML with the goal to recreate the probability distribution of some target data. Examples such as the Boltzmann machine \cite{nguyen2017inverse} achieve this by minimizing the distance between the probabilistic ansatz, at the core of the machine, and the heuristic estimation of the target distribution given by some sample data \cite{bengio}. The distinguishability between probability distributions can be computed using the aforementioned KLD \cite{Amari2016}. The KLD can be generalized for quantum density matrices $\hat{\rho}$ in form of the von Neumann relative entropy \cite{vedral1997quantifying,kappen2020learning},
\begin{equation}
	\label{eq:rhoDKL}
	D_{KL}\left[\hat{\rho}_1 : \hat{\rho}_2\right] = \tr\left[\hat{\rho}_1\log(\hat{\rho}_1)\right] - \tr\left[\hat{\rho}_1\log(\hat{\rho}_2)\right] \; .
\end{equation}
The generalized KLD quantifies how distinguishable $\hat{\rho}_1$ is from $\hat{\rho}_2$. The thermal density matrix $\hat{\rho} = \frac{1}{\mathcal{Z}}e^{-\beta \hat{H}}$ is fully defined by the system Hamiltonian $\hat{H}$ (and inverse temperature $\beta$). Here $\mathcal{Z}={\rm tr}[e^{-\beta \hat{H}}]$ is the partition function of $\hat{H}$. Thus, the thermal density matrix can be seen as a proxy for its defining Hamiltonian, and the KLD as a measure of distinguishability between two Hamiltonians. To emphasize this we denote the KLD for two thermal density matrices $\hat{\rho}_1 = \frac{1}{\mathcal{Z}_1}e^{-\beta \hat{H}_1}$ and $\hat{\rho_2} = \frac{1}{\mathcal{Z}_2}e^{-\beta \hat{H}_2}$ as $D_{KL}[\hat{H}_1:\hat{H}_2]$.

Given the target Hamiltonian $\hat{H}_{\rm{bare}}$, we seek to optimize the simpler effective model 
\begin{equation}
\label{eq:ham_parameterization}
\hat{H}_{\rm{eff}}(\pmb{\theta} ) = \sum_i \theta_i \hat{h}_i \;,
\end{equation}
by minimizing $D_{KL}[\hat{H}_{\rm bare}:\hat{H}_{\rm eff}(\pmb{\theta} )]$ with respect to the set of parameters $\{\theta_i\}$ corresponding to the operators $\{\hat{h}_i\}$. The minimization yields the vector of optimal couplings $\pmb{\theta}^*$ to represent $\hat{H}_{\rm{bare}}$ with $\hat{H}_{\rm{eff}}(\pmb{\theta}^* )$. In general this is an extremely challenging problem.

Specializing now to quantum impurity models of the type Eq.~\ref{eq:Hqim}, the KLD can be evaluated by representing the reduced thermal density matrix of the impurity after tracing out the bath, $\hat{\rho}_{\rm{imp}} = \tr_{\:\rm{bath}}\left[e^{-\beta \hat{H}}/\mathcal{Z}\right]$, as a \textit{classical} probability distribution. The fact that $\hat{\rho}_{\rm{imp}}$ can be viewed as a classical probability distribution can be seen as follows. Using the hybridization perturbation expansion \cite{haule2007quantum}, one can write the partition function $\mathcal{Z}_{\rm bare}$ of $\hat{H}_{\rm bare}$ as a sum over the weights $w_x$ of all impurity occupation diagrams, denoted $x$. We further subdivide each such diagram $x$ into a sum over all corresponding diagrams $w(\lbrace \alpha \rbrace_x)$ labelled in terms of impurity eigenstates $\lbrace \alpha \rbrace_x$, 
\begin{align}
	\begin{gathered}
		\label{eq:ZexpansionFinal}
		\mathcal{Z} =\mathcal{Z}^{\rm bath} \sum_{x} \sum_{\lbrace\alpha\rbrace_x} 
		w(\lbrace \alpha \rbrace_x) \;,\\
		w(\lbrace \alpha \rbrace_x) = e^{-\langle \hat{H}^{\rm{imp}}\rangle_{\lbrace \alpha \rbrace_x}}
		\Lambda_{\lbrace \alpha \rbrace_x}\det(\Delta_x) \;,
	\end{gathered}
\end{align}
where $w(\lbrace \alpha \rbrace_x)$ is the weight of a distinct Feynman diagram labelled by the eigenstate diagram $\lbrace \alpha \rbrace_x$, $\Delta_{x}$ is the antiperiodic hybridization matrix \cite{haule2007quantum}, $\Lambda_{\lbrace \alpha \rbrace_x}$ the sequence of impurity operators comprising the diagram $\lbrace \alpha \rbrace_x$ while being projected onto the eigenbasis of $\hat{H}^{\rm{imp}}$ and $\langle \hat{H}^{\rm{imp}}\rangle_{\lbrace \alpha \rbrace_x}$ is the average value of the impurity Hamiltonian over the diagram $\lbrace \alpha \rbrace_x$. We have also defined $\mathcal{Z}^{\rm bath}$ as the partition function of the free (decoupled) bath $\hat{H}^{\rm bath}$. The details of the derivation and precise definitions of the terms appearing in Eq.~\ref{eq:ZexpansionFinal} are given in Appendix \ref{app:convexity}.

From \ceqn{eq:ZexpansionFinal} we extract the distribution 
$$
P(\lbrace \alpha \rbrace_x) = (\mathcal{Z}^{\rm{bath}}/\mathcal{Z})w(\lbrace \alpha \rbrace_x) \;,
$$
which can be interpreted as a classical probability distribution provided that $ w(\lbrace \alpha \rbrace_x) > 0$. This distribution acts as a proxy for the impurity density matrix and hence also for the impurity Hamiltonian. As with the classical Boltzmann machine, the probability distribution $P$ is in the form of an energy-based model, with the weights $w(\lbrace \alpha \rbrace_x)$ here distributed according to the impurity Hamiltonian, $\hat{H}^{\rm imp}$. We can therefore evaluate the KLD 
\begin{align}
	\begin{gathered}
		\label{eq:P_KLD} 
		D_{KL}[\hat{H}_{\rm bare} : \hat{H}_{\rm eff}(\pmb{\theta} )] = \\
		\sum_x\sum_{\lbrace \alpha \rbrace_x:~\rm ad} P_{\rm bare}(\lbrace \alpha \rbrace_x) \log \left[ 
		\frac{P_{\rm bare}(\lbrace \alpha \rbrace_x)}{P_{\rm eff}(\lbrace \alpha \rbrace_x)} \right] \; ,
	\end{gathered}
\end{align}
where we have used the term \textit{admissible} ($\rm ad$) to denote diagrams that involve eigenstates of $\hat{H}^{\rm imp}_{\rm bare}$ for which $\hat{H}^{\rm imp}_{\rm eff}$ has analogues eigenstates. This is important because the dimensionality of the impurity subspace is lower for the effective model than in the bare model, by construction.
% optimization and convexity
To minimize the distinguishability between $\hat{H}_{\text{bare}}$ and $\hat{H}_{\rm{eff}}(\pmb{\theta} )$ we use gradient descent (GD) methods \cite{bengio}. For this, we need the analytic gradient of the KLD,
\begin{equation}
	\begin{gathered}
		\label{eq:gd}
		\nabla_{\theta} D_{\text{KL}}[\hat{H}_{\rm bare}:\hat{H}_{\rm eff}(\pmb{\theta} )] = 	\\
		\beta\langle \hat{\Omega}_{\rm ad}^\dagger\nabla_{\theta}\hat{H}_{\rm{eff}}\hat{\Omega}_{\rm ad}\rangle_{\rm{bare}}-\beta\langle \nabla_{\theta}\hat{H}_{\rm{eff}}\rangle_{\rm{eff}}\; ,
	\end{gathered}
\end{equation}
where the \textit{admissibility operator} $\hat{\Omega}_{\rm ad}$ connects the effective and bare Fock space, by mapping the effective eigenstates to analogous bare eigenstates (see \csec{sec:ad} for a detailed derivation and discussion). Thus, $\hat{\Omega}_{\rm ad}$ eliminates all non-admissible states from the bare thermal average such that $\langle \hat{\Omega}^\dagger_{\rm ad}\hat{\Omega}_{\rm ad}\rangle_{\rm{bare}}<1$, whereas on the effective Hilbert space it holds that,
\begin{equation}
	\label{eq:ad_effective}
	\langle \hat{\Omega}_{\rm ad} \hat{\Omega}^\dagger_{\rm ad}\rangle_{\rm{eff}} = \langle \hat{\mathbbm{1}}\rangle_{\rm{eff}} = 1 \;.
\end{equation}
From \ceqn{eq:gd} it follows immediately that the minimum is found when the impurity observables match, 
\begin{equation}
	\label{eq:gml_eff_bare_obs}
	\langle \hat{\Omega}_{\rm ad}^\dagger\hat{h}_i\hat{\Omega}_{\rm ad}\rangle_{\rm{bare}} = \langle \hat{h}_i\rangle_{\rm eff}
\end{equation}
for all effective impurity operators $\hat{h}_i$ of the effective model (from now on we simplify the notation $\langle \hat{\Omega}_{\rm ad}^\dagger\hat{h}_i\hat{\Omega}_{\rm ad}\rangle_{\rm{bare}} \rightarrow \langle \hat{h}_i\rangle_{\rm bare}$). Thus, in the precise sense of \ceqn{eq:P_KLD}, the optimal low-energy effective model $\hat{H}_{\eff}$ for a given bare model $\hat{H}_{\rm bare}$ is the one that matches the thermodynamic expectation values for all effective impurity operators $\hat{h}_i$ in the bare model. We emphasize that the KLD Eq.~\ref{eq:P_KLD} does not itself ever have to be evaluated. For the GD optimization, it is sufficient to evaluate the gradient of the KLD, which can be expressed in terms of the physical and readily computable observables in Eq.~\ref{eq:gd}. We refer to the process of optimizing an effective model to best represent the low-energy physics of our microscopic bare Hamiltonian by matching these key observables as \textit{`unsupervised model learning'} (UML).

A particularly attractive feature of UML for quantum impurity systems is the efficiency of the optimization problem. This follows from the fact that the second-order derivative of the KLD corresponds to the second-order derivative of the effective free energy $\mathcal{F}_{\rm eff} (\pmb{\theta}) = -1/\beta \log\tr\left[ \exp ( -\beta\hat{H}_{\rm eff}(\pmb{\theta}) )  \right]$. 
Therefore, provided all operators $\{\hat{h}_i\}$ mutually commute, the KLD is convex -- see \capp{app:convexity}. Thus we have a single, global minimum, which a gradient descent algorithm is guaranteed to find, making this a trivial optimization problem. We note that the gradient obtained from the KLD, \ceqn{eq:gd}, is equivalent to the gradient obtained in the \textit{maximum entropy principle} that was recently applied to the problem of Hamiltonian tomography in Ref.~\cite{anshu2021sample}.

The optimization problem is efficient, given the convexity of \ceqn{eq:P_KLD}. The computational cost of the optimization is therefore mainly controlled by the method used to compute the observables comprising the gradient in \ceqn{eq:gd}. The bottleneck is then the calculation of the observables $\langle \hat{h}_i\rangle_{\rm bare}$ in the bare model. However, this needs to be done just once (e.g.~with quantum Monte Carlo (QMC) methods \cite{li2020diagrammatic,SETH2016274}). On the other hand, although multiple calculations of $\langle \hat{h}_i\rangle_{\rm eff}$ are required during the optimization, these are performed on the simplified effective model and are therefore by construction inexpensive (they can be done with e.g.~the numerical renormalization group (NRG) method \cite{KWW,bulla2008numerical,mitchell2014generalized,*stadler2016interleaved}). For the results presented in the following, we have used continuous-time QMC (CT-QMC) and NRG to calculate the observables and the Adagrad or Adam GD methods \cite{duchi2011adaptive}.

%%%%%%%%%%%%%%%%%%
%%%%%%%%%%%%%%%%%%

\section{Designing an effective model}
\label{sec:gml_eff_ham}
The UML methodology described above makes no assumptions about the form of the bare impurity Hamiltonian (for example it could be an \textit{ab-initio} type model or in tight-binding form). However, for illustrative purposes, and without loss of generality, we assume here that the bare impurity Hamiltonian is akin to a \textit{Pariser-Parr-Pople} (extended Hubbard) model \cite{linderberg1968derivation},
\begin{equation}
	\label{eq:_PPP_}
	\hat{H}_{\rm bare}^{\rm imp} = \sum_{i,j,\sigma} \lambda_{ij}\hat{d}^{\dagger}_{i\sigma}\hat{d}^{\phantom{\dagger}}_{j\sigma} + \sum_{i} U_{i} \hat{n}_{i\uparrow}\hat{n}_{i\doa} + \sum_{i > j}U'_{ij} \hat{n}_{i}\hat{n}_{j} \; ,
\end{equation}
where $\hat{d}^{\dagger}_{i\sigma}$ ($\hat{d}_{i\sigma}$) creates (annihilates) a spin-$\sigma$ electron in impurity orbital $i$, whereas $\hat{n}_{i\sigma}=\hat{d}^{\dagger}_{i\sigma}\hat{d}^{\phantom{\dagger}}_{i\sigma}$ and $\hat{n}_{i}=\sum_{\sigma}\hat{n}_{i\sigma}$ are number operators. 
This form of $\hat{H}_{\rm bare}^{\rm imp}$ is particularly suited to describe molecules in the context of single-molecule transistors. In the following, we use the term `molecule' to describe the impurity degrees of freedom, as this emphasizes the important role of orbital complexity in the bare model. The total thermodynamic average charge on the molecule, $\langle\hat{Q}\rangle \equiv Q$,
can be controlled externally by means of a gate voltage $\hat{H}^{\rm gate} = V_{\rm g} \hat{Q}$, where $\hat{Q}=\sum_i \hat{n}_i$. 

In practice, the molecule degrees of freedom can be coupled to several non-interacting fermionic baths (metallic leads). For simplicity and concreteness, we now consider one single bath with particle-hole symmetry, $\hat{H}^{\rm bath} =\sum_{k,\sigma} \epsilon_{k}^{\phantom{\dagger}} \hat{c}_{ k \sigma}^{\dagger} \hat{c}_{ k \sigma}^{\phantom{\dagger}}$. We omit the `bare' label here because the bath will be common to both bare and effective models. Without any loss of generality, we cast this in the form of a 1d tight-binding chain,
\begin{equation}
\label{eq:bath}
\hat{H}^{\rm bath}  = \sum^\infty_{n = 0}\sum_\sigma t_n^{\phantom{\dagger}} (\hat{c}^\dagger_{n\sigma}\hat{c}_{n+1 \sigma}^{\phantom{\dagger}} + \hat{c}^\dagger_{n+1\sigma}\hat{c}_{n \sigma}^{\phantom{\dagger}} ) \;,
\end{equation}
where the chain parameters $\{t_n\}$ encode the bath density of states. This tridiagonal form can be derived explicitly using the Lanczos method \cite{allerdt2019numerically}. Onsite potentials are zero here due to particle-hole symmetry, but may be reinstated if needed without affecting the following discussions. For the RG analysis below, we note that the  \textit{Wilson chain} \cite{KWW} takes exactly the same form as Eq.~\ref{eq:bath}, but with chain parameters behave as $t_n \sim D \Lambda^{-n/2}$ for large $n$, where $D$ is the bare conduction electron half-bandwidth, and $\Lambda>1$ is the logarithmic discretization parameter. In that case we replace $\hat{H}^{\rm bath} \rightarrow \hat{H}^{\rm bath}_{\rm disc}$. 

The impurity degrees of freedom couple to bath at the end of the chain via the hybridization term,
\begin{equation}\label{eq:Hhybbare}
\hat{H}_{\rm bare}^{\rm hyb}=\sum_{i,\sigma} V_i^{\phantom{\dagger}}( \hat{d}^{\dagger}_{i\sigma} \hat{c}_{ 0 \sigma}^{\phantom{\dagger}}  + \hat{c}^{\dagger}_{ 0 \sigma}\hat{d}^{\phantom{\dagger}}_{i\sigma} ) \;.
\end{equation}
We denote the collection of all bare Hamiltonian couplings using the vector, 
$$
\vek{C} \equiv \lbrace \lambda_{11},\lambda_{12},.., U_1,U_2,.., U'_{11},U'_{12},.., V_{1},V_{2},..,V_{g}\rbrace \;.
$$

%%%%
% Effective Hamiltonian
By varying $V_g$ the ground state of the `molecule' can be tuned to a specific charge sector, which may be spin-degenerate. Of particular interest is the case where the molecule hosts an odd number of electrons and the ground state spin is $S=\tfrac{1}{2}$, although the arguments are general. With $E_{GS}$ the energy of the (possibly degenerate) ground state and $E_{\rm ex ,1}$ the energy of the first excited state, 
the active part of the molecule Fock space $\mathcal{H}_{\rm bare}^{\rm imp}$ at low temperatures $T \ll E_{\rm ex, 1} - E_{GS}$ is dominated by the Fock space of the ground state manifold. We can then map the full impurity Fock space of our bare model to an effective one spanning only the ground state manifold,
\begin{align}
	\mathcal{H}_{\rm bare}^{\rm imp} \rightarrow \mathcal{H}_{\rm eff}^{\rm imp} \;.
\end{align}
In terms of effective Hamiltonians, the above argument implies that we can map $\hat{H}_{\rm bare}^{\rm imp}+\hat{H}_{\rm bare}^{\rm hyb} \rightarrow \hat{H}_{\rm eff}^{\rm imp}+\hat{H}_{\rm eff}^{\rm hyb}$, with the bath Hamiltonian left unchanged. In particular, for a spin-degenerate ground state, the entire molecule can be represented by a single spin $\hat{\mathbf{S}}$ degree of freedom, and the molecule-bath hybridization term becomes a spin-exchange interaction. This mapping constitutes an immense reduction in complexity: for a molecule with $N$ orbitals, the dimension of the impurity Fock space is reduced from $4^N$ to $2S+1$. In the following, we discuss the details of how this mapping is achieved in practice, and provide a simple example demonstration.

%%%

For some parameterized effective model $\hat{H}_{\rm{eff}}(\pmb{\theta})$ we seek to optimize the coupling constants $\pmb{\theta}$ so that properties of the bare model are faithfully reproduced at low energies (as in Fig.~\ref{fig:schematic}). A crucial aspect of the GD optimization of \ceqn{eq:P_KLD} is to compare bare and effective physical observables. This must be done with care, since bare and effective models live in different Fock spaces and are defined at different energy scales. In the case of dynamical quantities such as Green's functions, a meaningful comparison of renormalized correlation functions can be achieved by rescaling procedures \cite{dolan1994geometrical}. But for static quantities, like the thermodynamic expectation values we are comparing, subtleties arise. In particular, a map between bare and effective eigenstates given by the admissibility operator $\hat{\Omega}_{\rm ad}$ is required. 

%%%%%%%%%%%%%%%

\subsection{Admissibility}\label{sec:ad}

Effective model operators are only defined on the effective model Fock space $\mathcal{H}_{\rm eff}^{\rm imp}$ and thus, to apply them to a state in the bare Fock space $\mathcal{H}_{\rm bare}^{\rm imp}$, we must project the states using the admissibility operator. One approach to construct the admissibility operator is to label bare and effective eigenstates with quantum numbers (QNs) corresponding to the symmetries that are common to both models,
\begin{align*}
	\hat{H}^{\rm imp} \ket{n} &= E_n \ket{n} \mapsto \hat{H}^{\rm imp} \ket{\vek{Q},\vek{q};  m_{\vek{Q},\vek{q}}} \\
	&= E_{ \vek{Q},\vek{q};m_{\vek{Q},\vek{q}}}\ket{\vek{Q},\vek{q} ; m_{\vek{Q},\vek{q}}}  \; ,
\end{align*}
where $\vek{Q} = (Q_1,Q_2,\dots ) $ denotes a vector of non-abelian QNs and $\vek{q} = (q_1,q_2,\dots )$ is a vector of abelian QNs. 
The label $m_{\vek{Q},\vek{q}} \in \lbrace 1,2,\dots,M_{\vek{Q},\vek{q}} \rbrace$ denotes an index which distinguishes multiplets with the same set of \qn{}s. $M_{\vek{Q},\vek{q}}$ is the number of such multiplets in a given sector, which is in general smaller in the effective model than in the bare model $M_{\rm eff} \le M_{\rm bare}$. This is required in order to have a meaningful reduction in complexity. 

By labelling our states with these \qn{}s we can identify which bare states are \textit{admissible} (``ad'' for short) on the level of symmetries. Importantly, we can remove \textit{inadmissible} states in the bare model that have no analogue in the effective model. For example, the full molecule Fock space of the bare model may contain high-lying spin $S=1$ states, whereas the effective model retains only the ground states with spin $S=\tfrac{1}{2}$. Therefore in this case only $S=\tfrac{1}{2}$ states of the bare model would be admissible. We regard any state that is labelled with a \qn{} combination that exists in the effective model Fock space as an admissible state of the bare model. 

For illustration, we now assume that $M_{\rm eff} = 1$. Then the admissibility operator can be written as a projector,
\begin{align}
	\label{eq:Pad}
	\hat{\Omega}_{\rm ad} = \sum_{\vek{Q},\vek{q}: \rm ad}\Bigg[ \ket{ \vek{Q},\vek{q}}_{\rm eff} \times\sum^{M_{\rm bare}}_{m_{\vek{Q},\vek{q}}=1}\bra{\vek{Q},\vek{q};m_{\vek{Q},\vek{q}}}_{\rm bare} \Bigg]\; ,
\end{align} 
where the notation `$\vek{Q},\vek{q}: \rm ad$' indicates that the sum runs only over the quantum numbers that label the eigenstates of $\hat{H}^{\rm imp}_{\rm eff}$. Thus, operators $\hat{h}_i$ of the effective model can be meaningfully computed in the bare Fock space, via,
\begin{equation}
	\chi_i = \langle \hat{\Omega}_{\rm ad}^\dagger\hat{h}_i\hat{\Omega}_{\rm ad}\rangle_{\rm{bare}} \; .
\end{equation}
We provide specific examples of this in action shortly.

% JBR: correct typo
There are exceptions where the reference-state quantum numbers differ in the bare and effective models and a little more care has to be taken to make them comparable. In particular, this applies to the definition of the charge quantum number, since the charge depends on the number of (occupied, fermionic) impurity degrees of freedom. But the number of degrees of freedom is smaller in the effective model than the bare one. For example, a molecule with $N$ orbitals at half-filling has a ground state charge $Q=N$. But the effective model may replace these degrees of freedom with a single local moment spin $S=1/2$, which can be regarded as a fermionic site subject to the constraint that the charge is $Q=1$. Indeed, the bare model may have $S=1/2$ representations in multiple charge sectors. One practical solution is to take only the charge sector $\tilde{Q}$ of the bare model which contains the lowest-energy $S=1/2$ states. Alternatively, one can sum over all charge sectors $\tilde{Q}$ that contain $S=1/2$ states. Different spin-$S$ multiplets with the same $\tilde{Q}$ in the bare model can also be summed over. 

%%%%%%%%%%%% 

\subsection{RG analysis}
The \gml{} approach yields the best couplings $\pmb{\theta}^*$ for the chosen Hamiltonian operators $\lbrace \hat{h}_i \rbrace$ in the effective model, \ceqn{eq:ham_parameterization}, given the target Hamiltonian $\hat{H}_{\rm{bare}}$. 
It may be advantageous to choose $\lbrace \hat{h}_i \rbrace$ in the most general and unbiased way possible, for maximum expressiveness and to facilitate automation of the process. However, we may also utilize physical insights in making this choice, to target the desired physics.

On the other hand, the RG procedure provides a systematic way to identify the structure of certain effective models. We briefly summarize the basics of the RG method below, as it will be useful in framing the following discussion. For a full description, see Ref.~\cite{kopietz2010introduction}.

The Wilsonian RG map \cite{wilson1971renormalization} $R_l$ for a Hamiltonian of the form $\hat{H}(\vek{C}) = \sum_i \theta_i \hat{h}_i$ (Eq.~\ref{eq:ham_parameterization}), with couplings $\vek{C} = (\theta_1,\theta_2,\dots)$ and associated interaction terms $\lbrace \hat{h}_i \rbrace$ allows one to eliminate correlations between states above and below a running energy scale cutoff $\exp(-l)D < D$, where $D$ is the bare model cutoff and $l > 1$. In the context of impurity models, $D$ is the conduction electron half-bandwidth. As the map $R_l$ is successively applied, we therefore focus on progressively lower and lower energy scales. The RG map produces a new Hamiltonian that is characterized by \textit{renormalized} couplings $\vek{C}'$,
\begin{align}
	R_l\left[ \hat{H}(\vek{C})  \right] = \hat{H}(\vek{C}') \; ,
\end{align}
while leaving the free energy invariant 
\begin{equation}
\label{eq:invariantF}
\mathcal{F}(\vek{C}) = \mathcal{F}(\vek{C}') \;.
\end{equation}
Within this framework, all terms $\hat{h}_i$ consistent with the symmetries and Fock space of the system are allowed to appear in $\hat{H}(\vek{C})$. For simplicity we assume that the set of interactions $\lbrace \hat{h}_i \rbrace$ is finite and does not change during the RG procedure. The continuous flow of the couplings $\vek{C}$ is described by the infinitesimal RG-transformation
\begin{equation}
	\label{eq:gellmann}
	\frac{d \mathbf{C}}{d l} = \lim_{\delta l \rightarrow 0} \frac{1}{\delta l} [R_{l+\delta l}(\mathbf{C}) - \mathbf{C}] \equiv \beta_l(\mathbf{C})  \; ,
\end{equation}
with $\beta_l$ being the Gell-Mann--Low  $\beta$-function \cite{kopietz2010introduction}.

An important property of RG is the existence of fixed points (FPs), which are characterized by a set of couplings $\vek{C}^*$ that are invariant under the RG transformation,
\begin{align}
	R_l\left[ \hat{H}(\vek{C}^*)  \right] = \hat{H}(\vek{C}^*) \equiv \hat{H}^* \; ,
\end{align}
and we call $\hat{H}^*$ the FP Hamiltonian. The RG transformation can be linearised near the FP Hamiltonian,
\begin{equation}
\label{eq:the_linearlised_RG}
R_l[\hat{H}^* + \delta \hat{H}] = \hat{H}^* + \mathcal{R}^*_l \cdot \delta \hat{H} + \mathcal{O}(\delta \hat{H}^2)\;,
\end{equation}
where $\delta\hat{H}$ is a perturbation to the fixed point. The FP can be further analysed by finding the eigensystem of the linearized RG transformation, $\mathcal{R}^*_l \cdot \hat{O}^*_i = \lambda^*_i \hat{O}^*_i$. The eigenvalue $\lambda^*_i$ tells us whether a given operator $\hat{O}^*_i$ drives the system away from the FP (for $|\lambda^*_i| > 1$) or brings us closer to the FP (for $|\lambda^*_i| < 1$). Physical perturbations to the FP can be decomposed in terms of these operators, $\delta \hat{H} = \sum_i \alpha_i \hat{O}^*_i$. If there are finite contributions to $\delta \hat{H}$ from operators with $|\lambda^*_i| > 1$ the perturbation is said to be \textit{relevant}. Conversely if there are only contributions from operators with $|\lambda^*_i| < 1$ the perturbation is \textit{irrelevant}. 

Within this framework, we can identify a class of \textit{minimal} effective models. As in Ref.~\cite{rigo2020machine}, we define a minimal effective model here as one that retains only the most RG relevant corrections to the low-energy stable FP. Elaborated effective models can also be formulated by including leading RG irrelevant operators. By contrast, \textit{minimally-constrained} effective models contain all possible operators consistent with the symmetries of the bare model that live in the reduced Fock space. We consider these various scenarios below.

We emphasize that although the RG procedure described above provides useful information on the \textit{structure} of FP Hamiltonians and their leading corrections, the RG equations for interacting quantum impurity models cannot be solved analytically exactly. Perturbative RG methods give useful insights \cite{kugler2018multiloop} but cannot provide quantitatively accurate predictions for effective model parameters in versatile settings. Non-perturbative solutions such as the numerical renormalization group \cite{KWW} allow exact results to be extracted \cite{hewson2004renormalized} but are only viable for sufficiently simple bare models, where $\hat{H}_{\rm bare}^{\rm imp}$ can be diagonalized exactly. In this paper we are interested precisely in the class of bare models that are beyond the reach of a direct solution using NRG, or at low temperatures inaccessible to QMC.

%%%%%%%%%%%%%%

\subsection{Schrieffer-Wolff revisited}

A paradigmatic example for the derivation of an effective model is the mapping of the bare symmetric single Anderson impurity model (\aim{}) to the effective Kondo model. The former features a spinful fermionic interacting orbital $\hat{d}_{\sigma}$ on the impurity, whereas the latter involves only a spin-half impurity operator $\hat{\mathbf{S}}_d$. The mapping eliminates the impurity charge states with zero or two electrons, and thus reduces the 4-dimensional impurity Fock space of the \aim{} to the 2-dimensional Fock space of the Kondo impurity spin.

We take the \aim{} as our bare Hamiltonian,
\begin{equation}
\begin{split}
	\label{eq:aim}
	\hat{H}^{\rm \aim{}}_{\text{bare}} = &\hat{H}^{\text{bath}}_{\rm disc} + \tfrac{1}{2}U_d(\hat{n}^d -1)^2 \\
	&+ V\sum_{\sigma} ( \hat{d}_{\sigma}^{\dagger}\hat{c}_{0\sigma}^{\phantom{\dagger}}+
	\hat{c}_{0\sigma}^{\dagger}\hat{d}_{\sigma}^{\phantom{\dagger}} ) \;,
 \end{split}
 \end{equation}
 where $\hat{n}^d=\sum_{\sigma}\hat{d}_{\sigma}^{\dagger}\hat{d}_{\sigma}$. As defined, the model possesses particle-hole symmetry and is hence at half filling (the ground state has impurity charge $Q=1$).
For $U_d\gg \Gamma$, with $\Gamma=\pi V^2/2D$ the impurity-bath hybridization, local moment formation on the impurity is expected.
 Under these circumstances, second order perturbation theory in the impurity-bath coupling $V$ yields the Kondo model,
\begin{equation}
	\label{eq:kondo}
	\hat{H}^{\text{K}}_{\rm eff} =\hat{H}^{\text{bath}}_{\rm disc} + J \hat{\mathbf{S}}_d \cdot \hat{\mathbf{S}}_0 \;, 
\end{equation}
where $\hat{\mathbf{S}}_0=\tfrac{1}{2}\sum_{\sigma,\sigma'}\pmb{\sigma}_{\sigma\sigma'}c_{0\sigma}^{\dagger}c_{0\sigma'}^{\phantom{\dagger}}$ is the conduction electron spin density at the impurity position. This is the famous Schrieffer-Wolff transformation (SWT) \cite{PhysRev.149.491,hewson1993}. 
The Kondo model is as such the \textit{minimal} effective model for the \aim{}, in the sense defined in the previous section.

The second-order perturbative calculation yields an estimate for the spin-exchange coupling, $J_{\rm SW}=8V^2/U_d$, which is naturally a rather crude approximation at finite $U_d/\Gamma$. Indeed, as discussed in Ref.~\cite{rigo2020machine}, physical properties such as the thermodynamic entropy and Kondo temperature of the effective model Eq.~\ref{eq:kondo} computed with $J_{\rm SW}$ do not match well those of the bare \aim{} Eq.~\ref{eq:aim} unless $U_d \gg D$. In fact, even when the SWT is carried out to infinite order in the impurity-bath hybridization \cite{chan2004exact}, the predicted value of $J$ still does not give accurate results for low-energy properties \cite{rigo2020machine}. The reason for this is well known: the SWT does not account for the bandwidth renormalization $D\to D_{\rm eff}$. In the small $U_d/D$ limit, Haldane found a perturbative estimate for this effect \cite{Haldane_1978}, with the intuitive result that $D_{\rm eff} \sim U_d$. This was put on a more concrete footing by Krishnamurthy \textit{et al} in Ref.~\cite{KWW} using RG arguments. Only recently in Ref.~\cite{rigo2020machine} was the precise form determined numerically by non-perturbative methods for all $U_d$ and $V$. The conclusion is that even with this highly simplified minimal effective model, the free energy and hence low-temperature thermodynamics can be faithfully reproduced -- provided the correct effective coupling $J$ is found.

%%%%%%%%%

\subsection{$\mathcal{F}$-learning for the Kondo model}\label{sec:F}

Since the \aim{} and the Kondo model are connected by the RG flow of the \aim{}, the low-temperature thermodynamics of the \aim{}, such as the impurity entropy flow, magnetic susceptibility, and the Kondo temperature $T_K$, can be exactly reproduced in an effective Kondo model with a properly chosen coupling constant $J$, provided $T_K\ll U_d$. As shown in Ref.~\cite{rigo2020machine}, this `true' $J$ can be found numerically exactly by solving the \aim{} and Kondo models non-perturbatively using e.g.~NRG, and matching their free energies
\begin{equation}
\mathcal{F}^{\rm eff}(J) = \mathcal{F}^{\rm bare}(U_d,V) \;, 
\end{equation}
which according to \ceqn{eq:invariantF} also matches their RG flows. 
This should be done at sufficiently low temperatures $T\ll U_d$ where the RG flows of the two models are expected to coincide. We refer to this model learning technique as $\mathcal{F}$-learning. This is a powerful method if one is interested in reproducing the low-temperature thermodynamics or low-frequency dynamical correlation functions of some bare model using a simpler effective model -- provided that both bare and effective models can be solved accurately down to a temperature $T\ll U_d$. However, for complex impurity problems, such a direct solution of the bare model might not be feasible. Furthermore, a coinciding free energy in bare and effective models does \textit{not} guarantee that all local thermodynamic observables will agree, as discussed in the next subsection. If such observables are the desired outcome from an effective model, $\mathcal{F}$-learning alone is not the best choice.

%%%%%%%%%

\subsection{Renormalization of observables}
Thermodynamic observables may themselves \textit{flow} under RG, and the full path taken, from high to low energies, affects the result. Since the effective model is defined on an energy scale below some cutoff, it cannot capture renormalization effects in the bare model at energies above this cutoff. The important implication of this is that we cannot expect even a optimal minimal effective model to capture simultaneously low-energy scales such as the Kondo temperature $T_K$, as well as the behavior of thermal expectation values. The former is controlled by the low-temperature free energy, whereas the latter depends on the entire RG flow.

\begin{figure}[t]
	\centering
	\includegraphics[width=0.5\textwidth]{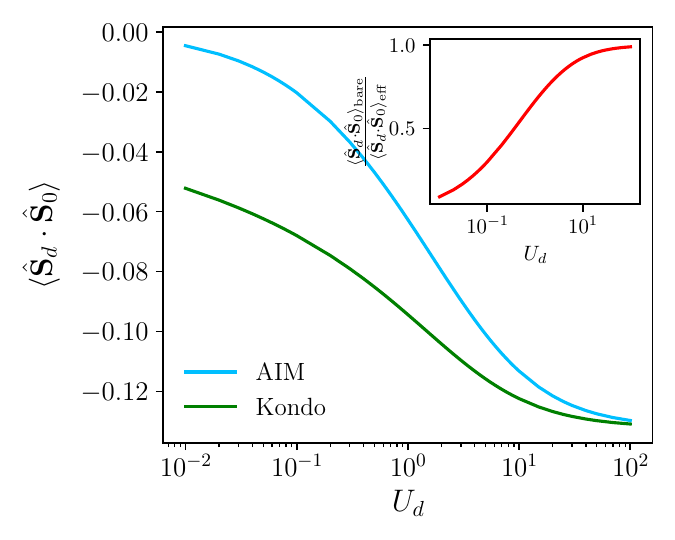}
\caption{Impurity-bath spin-spin expectation value ${\langle\hat{\mathbf{S}}_d \cdot \hat{\mathbf{S}}_0\rangle}$ computed with NRG at $T=0$ for the \aim{} \ceqn{eq:aim} (blue line) compared with the corresponding value in the Kondo model \ceqn{eq:kondo}, optimized by $\mathcal{F}$-learning (green). The inset shows their ratio, which captures the different renormalizations along the two paths to the same low-energy fixed point. Plotted as a function of $U_d$ for $8V^2/U_d=0.3$ and $D=1$.}
	\label{fig:gbf_observable_evolution}
\end{figure}

We demonstrate this point explicitly for the simplest example of the mapping from the \aim{} to the Kondo model. Using $\mathcal{F}$-learning \cite{rigo2020machine} we extracted the effective $J$ of the Kondo model that reproduces exactly the low-temperature thermodynamics and Kondo temperature $T_K$, for reference \aim{} systems with different bare $U_d$ (keeping $V^2/U_d$ constant and setting $D\equiv 1$). For each bare \aim{} and corresponding $\mathcal{F}$-optimized Kondo model, we computed with NRG the zero-temperature expectation value of the spin-spin correlator $\langle \hat{\mathbf{S}}_d\cdot \hat{\mathbf{S}}_0\rangle$. The results are shown in the main panel of Fig.~\ref{fig:gbf_observable_evolution}, demonstrating markedly different renormalization of this observable in bare and effective models, especially at small $U_d$. The ratio of the observables calculated in bare and effective models, shown in the inset, characterizes the different degree of renormalization in the two models along the RG flow. Only for $U_d \gtrsim 100D$ do the observables agree; this is consistent with the fact that the perturbative SWT becomes asymptotically exact in the large $U_d/\Gamma$ limit.

%%%%%%%%%%%%%%%%%%%%%%%%%%%%%%%%%%%%%%%%%
%%%%%%%%%%%%%%%%%%%%%%%%%%%%%%%%%%%%%%%%%

One can understand this result using the framework of Wilsonian RG described above. 
It is instructive to interpret the RG procedure as a reparameterization of the free energy, $\mathcal{F}$. In this picture, the renormalization is achieved by replacing the original coordinate system $\vek{C}$ with a new coordinate system $\mathbf{C}'$ that leaves $\mathcal{F}$ invariant, as required by \ceqn{eq:invariantF}. Thus we have $\vek{C} \rightarrow  \mathbf{C}'(\vek{C})$  
and thermodynamic observables can be interpreted as covariant tensors that can be transformed as such 
between coordinate systems \cite{dolan1994geometrical}. For the effective model Eq.~\ref{eq:ham_parameterization}, we may therefore write,
\begin{equation}
	\label{eq:floweq1}
	\begin{aligned}
		\langle \hat{h}_i \rangle_{\rm eff} &= \frac{\partial}{\partial \theta_i} \mathcal{F}_{\rm eff}(\pmb{\theta}) \\
		&= \sum_j \frac{\partial \theta^*_j }{\partial \theta_i}\frac{\partial}{\partial \theta^*_j} \mathcal{F}_{\rm eff}(\pmb{\theta}^*)= \sum_j \chi^*_{\theta_j} \frac{\partial \theta^*_j }{\partial \theta_i} \; ,
	\end{aligned}
\end{equation}
with $\pmb{\theta}^*$  being the renormalized effective model parameters at the low-energy \fp{}, and where $\chi^*_{\theta_j}$ is the expectation value of the operator $\hat{h}_j$ evaluated in the FP Hamiltonian. The matrix of derivatives $\frac{\partial \theta^*_j }{\partial \theta_i}$ contains information about the RG flow to the FP, and can in principle be obtained from the $\beta$-function, \ceqn{eq:gellmann}. With this information we can make a connection between the renormalized value of the observable and its \fp{} value. The desired observable is found to be a linear combination of all possible \fp{} expectation values.

\begin{sloppypar}
Taking again the example of the \aim{} to Kondo model mapping, the low-energy stable FP (of both models) is of course the strong coupling (SC) FP \cite{wilson1971renormalization,KWW}. In the effective Kondo model, the SC FP Hamiltonian is characterized by $J^*\to \infty$.
Therefore ${\chi^*_J\equiv \partial_{J^*} \mathcal{F}_{\rm eff}(J^*) = {\langle\hat{\mathbf{S}}_d \cdot \hat{\mathbf{S}}_0\rangle}^*}$ evaluated at the SC FP takes the value,
\begin{equation}
\chi^*_J = -3/4\;,
\end{equation}
which as such embodies the spin-singlet formed between the impurity and bath. However the actual value of the correlator as measured in the Kondo model, ${\chi_J\equiv \partial_J \mathcal{F}_{\rm eff}(J)}$, carries information about the RG flow, 
\begin{equation}
\langle\hat{\mathbf{S}}_d \cdot \hat{\mathbf{S}}_0\rangle =  -\frac{3}{4} \frac{\partial J^*}{\partial J} \;.
\end{equation}
It is tempting to use perturbative scaling techniques \cite{Anderson_1970} to evaluate the nontrivial factor $\tfrac{\partial J^*}{\partial J}$, but we find that these yield rather poor approximations to the true value of the correlator obtained by NRG.
\end{sloppypar}

To understand the flow of the same observables $\langle \hat{h}_i \rangle$ in the bare model, we have to add $\hat{h}_i$ to the bare model as a source term. This is because the effective interactions $\hat{h}_i$ do not typically appear in the original bare model. Thus we may write,
\begin{equation}
	\begin{aligned}
		\label{eq:floweq2}
		\langle \hat{h}_i \rangle_{\rm bare} &= \frac{\partial}{\partial \theta_i} \mathcal{F}_{\rm bare}(\vek{C}; \theta_i) \Big\vert_{\theta_i = 0} \\
		&= \left[\frac{\partial \theta^*_i }{\partial \theta_i}\frac{\partial}{\partial \theta^*_i} +\sum_j\frac{\partial C^*_j }{\partial \theta_i}\frac{\partial}{\partial C^*_j}\right] \mathcal{F}_{\rm bare}(\vek{C}^*; \theta^*_i) \Big\rvert_{\theta_i = 0} \\
		&= \chi^*_{\theta_i}\frac{\partial \theta^*_i }{\partial \theta_i}\Big\vert_{\theta_i = 0} + \sum_j\chi^*_{C_j}\frac{\partial C^*_j }{\partial \theta_i}\Big\vert_{\theta_i = 0} \; ,
	\end{aligned}
\end{equation}
where $\vek{C}^*$ are the bare model \fp{} parameters, and $\chi^*_{\theta_i}$ or $\chi^*_{C_j}$ are expectation values evaluated in the \fp{} Hamiltonian of the bare model.

In the low-energy limit, both bare and effective models share the same FP, by construction. Therefore in the bare model we may write the FP parameters $\theta_j^*\equiv \theta^*_j({\vek{C}})$ as functions of the bare model parameters. From this it follows that 
\begin{align}
	\langle \hat{h}_i \rangle_{\rm bare} =\sum_j \chi^*_{\theta_j} \frac{\partial \theta^*_j }{\partial \theta_i}\Big\vert_{\theta_i = 0} \;,
\end{align}
where $\chi_{\theta_j^*}$ are the FP expectation values of the effective model operators $\hat{h}_j$. Since the FP is the same, these expectation values are the same in bare and effective models.
For the \aim{} to Kondo mapping, we find,
\begin{equation}
	\langle\hat{\mathbf{S}}_d \cdot \hat{\mathbf{S}}_0\rangle_{\rm eff} \le \langle\hat{\mathbf{S}}_d \cdot \hat{\mathbf{S}}_0\rangle_{\rm bare} \;,
\end{equation}
as demonstrated numerically in Fig.~\ref{fig:gbf_observable_evolution}. 
This implies that,
\begin{equation}
	\left[\frac{\partial J^*}{\partial J}\right]_{\rm bare} \Big/ \left[\frac{\partial J^*}{\partial J}\Big\vert_{J = 0}  \right]_{\rm eff}\le 1 \;,
\end{equation}
where equality is only approached in the large $U_d$ limit of the \aim{} where the RG flow is equivalent to that of the Kondo model.

The above has consequences for automated model learning. When optimizing an effective model with UML using GD, Eq.~\ref{eq:gd}, we should match the full set of observables $\langle \hat{h}_i \rangle_{\rm bare} = \langle \hat{h}_i \rangle_{\rm eff}$ for all effective model operators $\hat{h}_i$.
This in turn implies that we enforce,
\begin{equation}
\label{eq:relate_flows}
\Bigg[\frac{\partial \theta^*_j }{\partial \theta_i}\Bigg]_{\rm eff} = \Bigg[\frac{\partial \theta^*_j }{\partial \theta_i}\Big\vert_{\theta_i = 0}\Bigg]_{\rm bare} \;.
\end{equation}
However, the RG flow to reach the low-energy FP is not the same in bare and effective models since the latter typically has fewer FPs. Therefore enforcing Eq.~\ref{eq:relate_flows} will give a solution for which the free energy of the effective model is in general not the same as that of the bare model. The low-temperature thermodynamic entropy, Kondo scales etc, are therefore only in good agreement if the matched observables do not flow \textit{strongly}. 
For the Kondo model, we see that $\langle \hat{\mathbf{S}}_d\cdot \hat{\mathbf{S}}_0\rangle$ is in general renormalized rather differently compared with the \aim{}.

Our conclusion is that the KLD Eq.~\ref{eq:P_KLD} is only a good measure of Hamiltonian distinguishability if the bare and effective models have similar RG flows. A \textit{minimal} effective model may therefore be poorly suited to this kind of optimization.

The question addressed in the next section is: can more general effective model structures in principle capture the Kondo scale and thermal expectation values simultaneously, at least approximately? If so, can UML be used to automate the optimization while remaining computationally tractable?

%%%%%%%%%%

\section{Minimally-constrained\\~~~~effective models}\label{sec:minconst}
We now go beyond minimal models, like the Kondo model, and introduce a family of minimally constrained (MC) effective models. With a given choice for the reduced Fock space of the effective model, MC models are characterized by a completely general operator structure, constrained only to respect the \textit{symmetries} of the bare model. This allows for the inclusion and parameterization of all possible RG relevant, marginal and irrelevant operators.  We show in the following sections that such an e ffective model can much more accurately capture the physics of the bare model. Because its RG flow is more comparable with the bare model, MC models are better able to capture both the behavior of local observables as well as the Kondo temperature and associated universal low-energy Kondo physics. This also means that the UML strategy can be used to find good effective models, when the starting point is an MC model.

To construct an MC model, we first must decide on the reduced Fock space in which the effective model lives. In this regard, we note that by replacing the impurity part of the bare Hamiltonian by operators spanning only the ground state manifold, we also implicitly change the nature of the impurity-bath hybridization term. Consider for example the classic mapping of the AIM to the Kondo model. The impurity degrees of freedom $\hat{d}_{\sigma}$ are mapped to a spin-$\tfrac{1}{2}$ operator $\hat{\mathbf{S}}_d$, but the impurity-bath tunneling $ V\sum_{\sigma} ( \hat{d}_{\sigma}^{\dagger}\hat{c}_{0\sigma}^{\phantom{\dagger}}+{\rm H.c.})$ in Eq.~\ref{eq:aim} is also replaced by a spin-spin exchange coupling $J \hat{\mathbf{S}}_d \cdot \hat{\mathbf{S}}_0$ in Eq.~\ref{eq:kondo}. 

An alternative way to formulate the mapping is to take an extended view on what constitutes the `impurity' to include also the local bath orbitals to which the bare impurity degrees of freedom couple. In so doing, we also redefine our bath to exclude these local orbitals. We thus repartition the bare model as 
$\hat{H}_{\rm bare}=\hat{\tilde{H}}^{\rm imp}_{\rm bare} +\hat{H}^{\rm bath}_1 + \hat{H}^{\rm hyb}_1$ where our extended impurity is now given by,
\begin{equation}\label{eq:ext_imp}
\hat{\tilde{H}}^{\rm imp}_{\rm bare}=\hat{H}_{\rm bare}^{\rm imp}+\hat{H}_{\rm bare}^{\rm hyb} \;.
\end{equation}
The extended impurity Fock space is thus,
\begin{equation}
\label{eq:newfockbare}
\tilde{\mathcal{H}}^{\rm imp}_{\rm bare}=\mathcal{H}^{\rm imp}_{\rm bare} \otimes \mathcal{H}^{\rm bath}_{\rm 0} \;.
\end{equation}
The bath and hybridization terms are correspondingly redefined,
\begin{align}
\hat{H}^{\rm bath}_N &=  \sum^\infty_{n = N}\sum_\sigma t_n^{\phantom{\dagger}} (\hat{c}^\dagger_{n\sigma}\hat{c}_{n+1 \sigma}^{\phantom{\dagger}} + \hat{c}^\dagger_{n+1\sigma}\hat{c}_{n \sigma}^{\phantom{\dagger}} ) \;, \label{eq:newbath}\\
\hat{H}^{\rm hyb}_N &= \sum_\sigma t_{N-1}^{\phantom{\dagger}} (\hat{c}^\dagger_{N-1\sigma}\hat{c}_{N \sigma}^{\phantom{\dagger}} + \hat{c}^\dagger_{N\sigma}\hat{c}_{N-1 \sigma}^{\phantom{\dagger}} ) \label{eq:newhyb} \;,
\end{align}
where $N=1$ indicates the new starting site index of the redefined bath chain.
This has the advantage that the new bath and hybridization terms are always the same in bare and effective models. We have emphasized this by now removing the `bare' label from these terms. 

The MC effective model is similarly structured,
\begin{equation}\label{eq:MCeff}
\hat{H}^{\rm MC}_{\rm eff}(\pmb{\theta}) = \hat{\tilde{H}}^{\rm imp}_{\rm eff}(\pmb{\theta})+\hat{H}^{\rm bath}_1+ \hat{H}^{\rm hyb}_1 \;,
\end{equation}
where, 
$\tilde{\tilde{H}}^{\rm imp}_{\rm eff}(\pmb{\theta})= \sum_i \theta_i \hat{h}_i$
and the effective impurity operators $\hat{h}_i$ live in the extended MC effective Fock space,
\begin{equation}
\label{eq:newfockeff}
\tilde{\mathcal{H}}^{\rm imp}_{\rm eff}=\mathcal{H}^{\rm imp}_{\rm eff} \otimes \mathcal{H}^{\rm bath}_{\rm 0} \;.
\end{equation}
As before, $\mathcal{H}^{\rm imp}_{\rm eff}$ spans the ground state manifold of $\hat{H}_{\rm bare}^{\rm imp}$ and $\mathcal{H}^{\rm bath}_{\rm 0}$ is the Fock space of the local bath orbital(s).

%%%%%%%%%%%%%%%%%%%%

\subsection{Corrections to the Kondo model}
As the simplest example, we again consider the particle-hole symmetric AIM as our bare model, Eq.~\ref{eq:aim}. The MC effective model can be found by following the above arguments. With a spin-$\tfrac{1}{2}$ operator $\hat{\mathbf{S}}_d$ representing the impurity and $\hat{c}_{0\sigma}$ for the local bath orbital, the effective impurity Fock space $\tilde{\mathcal{H}}^{\rm imp}_{\rm eff}$ is 8-dimensional. Since we have conserved spin $S$, spin projection $S_z$ and charge $Q$, we may label the basis states of the MC effective impurity Hamiltonian by these quantum numbers, $|Q,S;S_z\rangle$.
Then $\tilde{H}^{\rm imp}_{\rm eff}(\pmb{\theta})$ assumes a block diagonal structure, with each quantum number block in this case being simply a $1\times 1$ scalar. The allowed configurations are,
\begin{center}
\begin{tabular}{ |c|c|c| } 
 \hline
 $~~Q~~$ & $~~S~~$ & $~~~S_z~~~$ \\ 
 \hline \hline
$\pm 1$ & $\tfrac{1}{2}$ & $\pm\tfrac{1}{2}$ \\
 \hline
 0 & $0$ & $0$ \\
 \hline
 0 & $1$ & $0,~\pm 1$ \\
 \hline
\end{tabular}
\end{center}
where we specify the charge $Q$ with respect to the 2-electron half-filled state. Due to SU(2) spin symmetry, states with the same $Q$ and $S$ are degenerate; and due to particle-hole symmetry, states with $Q=\pm 1$ are degenerate. The most general MC effective Hamiltonian defined in this Fock space, consistent with these symmetries, can therefore be parameterized by just three distinct Hamiltonian matrix elements. Because in this case each quantum number block contains a unique basis state, the Hamiltonian operator can be entirely specified in terms of projectors onto these symmetry spaces, 
\begin{equation}\label{eq:Pqs}
\hat{P}_{QS}= \sum_{S_z}|Q,S;S_z\rangle\langle Q,S;S_z| \;,
\end{equation}
according to,
\begin{equation}\label{eq:Hprojfull}
\tilde{H}^{\rm imp}_{\rm eff}(\pmb{\theta}) = \theta_{\pm 1,1/2}(\hat{P}_{-1,1/2} + \hat{P}_{+1,1/2}) + \theta_{0,0}\hat{P}_{0,0} + \theta_{0,1}\hat{P}_{0,1} \;.
\end{equation}
These projectors can be written out explicitly as,
\begin{align}
	\label{eq:projectors}
	\hat{P}_{0,1} &= \ket{\uparrow,\Upa}\bra{\uparrow,\Upa}+\ket{\downarrow,\Doa}\bra{\downarrow,\Doa}+\frac{1}{2}(\ket{\uparrow,\Doa}\bra{\uparrow,\Doa}  \nonumber \\
	& + \ket{\downarrow,\Upa}\bra{\downarrow,\Upa} + \ket{\downarrow,\Upa}\bra{\uparrow,\Doa} + \ket{\uparrow,\Doa}\bra{\downarrow,\Upa}) \nonumber \\
	\hat{P}_{\pm 1,1/2} &= ( \ket{0}\bra{0} + \ket{\uparrow\downarrow}\bra{\uparrow\downarrow} )\otimes( \ket{\Upa}\bra{\Upa} + \ket{\Doa}\bra{\Doa} ) \nonumber\\
	\hat{P}_{0,0} &= \frac{1}{2}(\ket{\uparrow,\Doa}\bra{\uparrow,\Doa} + \ket{\downarrow,\Upa}\bra{\downarrow,\Upa}- \ket{\downarrow,\Upa}\bra{\uparrow,\Doa}  \nonumber \\ 
	&- \ket{\uparrow,\Doa}\bra{\downarrow,\Upa})  \; ,
\end{align}
where $|\phi,\sigma\rangle\equiv |\phi\rangle_0\otimes|\sigma\rangle_d$ with $\phi=\{0,\uparrow,\downarrow,\uparrow\downarrow\}$ for the local bath orbital and $\sigma=\{\Upa,\Doa\}$ for the impurity spin.

The identity operator can also be resolved in this basis,
\begin{equation}
\hat{\mathbbm{1}}_{\rm eff} = \sum_{Q,S} \hat{P}_{QS} \; ,
\end{equation}
such that $\langle \hat{\mathbbm{1}}_{\rm eff} \rangle_{\rm eff}=1$. We can therefore use the identity operator to eliminate one of the projectors in  \ceqn{eq:Hprojfull}. As an example we eliminate the $\hat{P}_{0,0}$ projector and write,
\begin{equation}\label{eq:Himp_eff_P}
\tilde{H}^{\rm imp}_{\rm eff}(\pmb{\theta}) = \theta_{\pm 1,1/2}(\hat{P}_{-1,1/2} + \hat{P}_{+1,1/2}) + \theta_{0,1}\hat{P}_{0,1} + \theta_\mathbbm{1} \hat{\mathbbm{1}}_{\rm eff}\;,
\end{equation}
where $\theta_\mathbbm{1}=\theta_{0,0}$ and the (as yet still unknown) parameters $\theta_{\pm 1,1/2}$ and $\theta_{0,1}$ have been suitably rescaled. This is useful because the identity operator strictly does not flow under RG. Thus, the term $\theta_\mathbbm{1} \hat{\mathbbm{1}}_{\rm eff}$ cannot change the RG flow of the effective model or affect thermal expectation values of the projectors $\langle \hat{P}_{QS}\rangle_{\rm eff}$ evaluated in the effective model. Further, setting $\theta_\mathbbm{1}=0$ corresponds to a uniform shift of all energy levels and does not affect the dynamics. For the symmetric AIM, the ground state lives in the $Q=0$, $S=0$ sector, so this choice is equivalent to measuring energies relative to the ground state energy. Importantly, the UML optimization, which is based on matching thermal expectation values, is sensitive to this elimination of a projector, as the unassuming operator $\hat{\mathbbm{1}}_{\rm eff}$ becomes non-trivial in the bare model  ${ \hat{\Omega}^\dagger_{\rm ad}\cdot\hat{\mathbbm{1}}_{\rm eff} \cdot \hat{\Omega}_{\rm ad}} \neq \hat{\mathbbm{1}}_{\rm bare}$. The intricacies that arise from this sensitivity are discussed in the detail in Sec.~\ref{sec:uml_opt_mo}.

In any case we are left with an MC effective model featuring just two tunable parameters. The physical interpretation of Eq.~\ref{eq:Himp_eff_P} is manifested by expressing the projectors in second quantized form,
\begin{align}
	\label{eq:1emolP}
	\hat{P}_{0,1} &= \frac{3}{4}(\hat{n}_{0\uparrow} + \hat{n}_{0\downarrow} - 2 \hat{n}_{0\uparrow} \hat{n}_{0\downarrow} ) + \hat{\mathbf{S}}_d \cdot \hat{\mathbf{S}}_0 \nonumber \\
	\hat{P}_{0,0} &= \frac{1}{4}(\hat{n}_{0\uparrow} + \hat{n}_{0\downarrow} - 2 \hat{n}_{0\uparrow} \hat{n}_{0\downarrow} ) - \hat{\mathbf{S}}_d \cdot \hat{\mathbf{S}}_0  \\
	\hat{P}_{\pm1,1/2} &= \hat{\mathbbm{1}}_{\rm eff} - \hat{n}_{0\uparrow} - \hat{n}_{0\downarrow} + 2 \hat{n}_{0\uparrow} \hat{n}_{0\downarrow} \nonumber  \; ,
\end{align}
where $\hat{n}_{0\sigma}=\hat{c}_{0\sigma}^{\dagger}\hat{c}_{0\sigma}^{\phantom{\dagger}}$. Recombining these projectors yields
a different parameterization of Eq.~\ref{eq:Himp_eff_P} (but with the same expressibility). We write it as,
\begin{equation}\label{eq:Himp_eff_JU}
\tilde{H}^{\rm imp}_{\rm eff}(\pmb{\theta}) \equiv \hat{H}^{JU_0}_{\text{eff}}= J \hat{\mathbf{S}}_d \cdot \hat{\mathbf{S}}_0 + \tfrac{1}{2}U_0 (\hat{n}_0 -\hat{\mathbbm{1}}_{\rm eff})^2 + L\hat{\mathbbm{1}}_{\rm eff} \; ,
\end{equation}
where $\hat{n}_0=\sum_\sigma \hat{n}_{0\sigma}$
and $J=\theta_{0,1}-\theta_{0,0}$, $U_{0}=2\theta_{\pm1,1/2}-\tfrac{3}{2}\theta_{0,1}-\tfrac{1}{2}\theta_{0,0}$, $L=\tfrac{3}{4}\theta_{0,1}+\tfrac{1}{4}\theta_{0,0}+\theta_\mathbbm{1}$. Again, the term involving $\hat{\mathbbm{1}}_{\rm eff}$ can be neglected and we set $L=0$. The effective model written in this way is equivalent to a Kondo model with an additional electron-electron interaction term on the local bath orbital.

Eq.~\ref{eq:Himp_eff_P}, and equivalently Eq.~\ref{eq:Himp_eff_JU}, are the most general forms of the effective model that live in the extended Fock space $\tilde{\mathcal{H}}^{\rm imp}_{\rm eff}$, consistent with SU(2) spin symmetry, U(1) charge conservation, and particle-hole symmetry.

Although the minimal effective model for the AIM is the Kondo model Eq.~\ref{eq:kondo}, the MC effective model $\hat{H}^{\rm MC}_{\text{eff}}(J,U_0)$, with $\tilde{H}^{\rm imp}_{\rm eff}$ parameterized according to Eq.~\ref{eq:Himp_eff_JU}, is precisely the Kondo model \textit{plus the leading RG-irrelevant perturbation} to the local moment fixed point identified by Krishnamurthy, Wilkins and Wilson in Ref.~\cite{KWW}. The extended-impurity approach thus allows us to define more general models that include such RG marginal and irrelevant terms, beyond those obtained by e.g.~a Schrieffer-Wolff transformation.

\rev{In terms of the interpretability of effective models learned by UML, we note that the symmetry analysis already offers insights into the kinds of coupling terms that can arise in a given extended impurity space. The fact that the physical impurity hosts a local moment is exploited directly to reduce the dimensionality of the active impurity space; this has implications for the interactions appearing in the effective model. By formulating the resulting symmetry projectors in second-quantized form, as per Eq.~\ref{eq:1emolP}, we obtain a more physical interpretation of these terms. Indeed, recombining these in the form of Eq.~\ref{eq:Himp_eff_JU} makes a connection to the RG theory of the low-energy properties of the bare model. However, the numerical values of the optimized coupling constants remain somewhat inscrutable. Only in perturbative limits do these effective parameters take a simple form in terms of the bare model parameters.}

When the AIM is the bare model, the extended impurity Fock space dimension is reduced modestly from 16 to 8, but the parametric complexity $(U_d,V)\to (J,U_0)$ is the same and we do not appear to have gained much. However, we emphasize that the AIM is \textit{not} the intended application of this method. Rather, we are interested in being able to systematically derive expressive effective models for complex impurity systems. We provide examples of this in the following. The AIM is however still a useful benchmark test case which we continue to explore.

%%%%%%%%%%

\subsection{UML optimization of the MC effective model}
\label{sec:uml_opt_mo}
With the form of the MC effective model now at hand, we turn to the problem of optimizing its parameters using the UML method. As explained in Sec.~\ref{sec:method}, this amounts to matching the observables $\langle \hat{\Omega}^{\dagger}_{\rm ad} \hat{h}_i^{\phantom{\dagger}} \hat{\Omega}^{\phantom{\dagger}}_{\rm ad}\rangle_{\rm bare}=\langle \hat{h}_i\rangle_{\rm eff}$ in bare and effective models, where $\hat{h}_i$ are the operators comprising the MC effective model, and $\hat{\Omega}_{\rm ad}$ is the admissibility operator that projects onto the reduced Fock space of the effective model. The explicit form of these will be discussed below, and an explicit example of the UML optimization given at the level of the AIM.

First, we note that for the MC effective model (in either form, Eq.~\ref{eq:Himp_eff_P} or Eq.~\ref{eq:Himp_eff_JU}), we have two observables to match and \textit{two} parameters to learn when comparing with the bare model. In particular, we emphasize that the identity $\hat{\mathbbm{1}}_{\rm eff}$ is \textit{not} a matchable observable, because the Fock spaces of bare and effective models are different by construction. Therefore although $\langle \hat{\mathbbm{1}}_{\rm eff} \rangle_{\rm eff} = 1$ in the effective model, we have $\langle \hat{\Omega}^\dagger_{\rm ad} \cdot \hat{\mathbbm{1}}_{\rm eff} \cdot \hat{\Omega}_{\rm ad} \rangle_{\rm bare} < 1$ in the bare model. From this it follows that it is in fact \textit{impossible} to demand that,
\begin{equation}
	\langle \hat{\Omega}^\dagger_{\rm ad} \cdot \hat{P}_{QS} \cdot \hat{\Omega}_{\rm ad} \rangle_{\rm bare} = \langle \hat{P}_{QS} \rangle_{\rm eff} \;,
\end{equation}
be satisfied simultaneously for the full set of projectors $\hat{P}_{QS}$. Importantly, we conclude that the optimization problem \ceqn{eq:P_KLD} is \textit{overdetermined} when using these projectors. Therefore at least one projector must be omitted from the UML observable matching process. In Eq.~\ref{eq:Himp_eff_P} we eliminated $\hat{P}_{0,0}$, but other choices can be made.   

Another key point is that matching different combinations of observables may yield different results for the optimized effective model parameters. The operator content and scaling dimensions for different observables means they are renormalized differently. Since bare and effective models have different RG flows in the UV, different effective model parameters are in general needed to ensure agreement between for a given set of observables. We return to the systematic selection of optimal observables below; but here we simply note that observables that flow the \textit{least} are the best choice.

Although the impurity Hamiltonian terms in $\hat{H}_{\rm eff}^{JU_0}$ in Eq.~\ref{eq:Himp_eff_JU} are arguably more physically meaningful, the decomposition in terms of symmetry projectors in Eq.~\ref{eq:Himp_eff_P} turns out to be most advantageous for UML. This is because (i) the projectors have a direct meaning in the bare model, since bare and effective models have the same \textit{symmetries}; (ii) the projectors mutually commute, and so the optimization problem is \textit{convex}; (iii) the action of the admissibility operator on the projectors is straightforward. 

%%%%%%

\subsubsection{UML for the AIM}
We first look at the AIM as our bare model. As explained above, we have two observables to match within the UML optimization procedure. Working with the projectors in Eq.~\ref{eq:Himp_eff_P} we tune the effective model parameters $\theta_{\pm1,1/2}$ and $\theta_{0,1}$ via GD according to Eq.~\ref{eq:gd}. Importantly, the KLD loss function itself need not be computed at any stage; the optimization can be done with a knowledge of its gradient, which involves only the observables $\langle \hat{P}_{\pm1,1/2}\rangle_{\rm eff}$ and $\langle \hat{P}_{0,1}\rangle_{\rm eff}$ in the effective model, and 
$\langle \hat{\Omega}^{\dagger}_{\rm ad}\hat{P}_{\pm1,1/2}\hat{\Omega}^{\phantom{\dagger}}_{\rm ad}\rangle_{\rm bare}$ and $\langle \hat{\Omega}^{\dagger}_{\rm ad} \hat{P}_{0,1}\hat{\Omega}^{\phantom{\dagger}}_{\rm ad}\rangle_{\rm bare}$ in the bare model. The optimal solution is found when both:
\begin{equation}\label{eq:aimobs}
\begin{split}
\langle \hat{P}_{\pm1,1/2}\rangle_{\rm eff} &= \langle \hat{\Omega}^{\dagger}_{\rm ad}\: \hat{P}_{\pm1,1/2} \: \hat{\Omega}^{\phantom{\dagger}}_{\rm ad}\rangle_{\rm bare} \;,\\
\langle \hat{P}_{0,1}\rangle_{\rm eff} &= \langle \hat{\Omega}^{\dagger}_{\rm ad} \: \hat{P}_{0,1} \:\hat{\Omega}^{\phantom{\dagger}}_{\rm ad}\rangle_{\rm bare} \;.
\end{split}
\end{equation}
The $Q$ and $S$ quantum numbers have physical meaning in the enlarged Fock space $\tilde{\mathcal{H}}_{\rm bare}^{\rm imp}$ of the bare model, and so $\hat{P}_{QS}$ can be defined in the bare model. The admissibility operator projects onto the $S=\tfrac{1}{2}$ states of the impurity. For the AIM, this means the two local moment impurity states with $n_d=1$. Because of the simplicity of the AIM and the symmetry constraints, we can construct $\hat{\Omega}^{\phantom{\dagger}}_{\rm ad}=\hat{n}_d-2\hat{n}_{d\uparrow}\hat{n}_{d\downarrow}$ and use the definitions of the projectors given already in Eq.~\ref{eq:1emolP}.

A key advantage of UML is that the optimization is done via physical quantities that can be obtained using any numerical method able to compute static thermodynamic observables, including \textit{ab initio} methods. Indeed, different methods can be used for bare and effective models, if desired. For the AIM considered in this section, we use NRG to compute the observables in both bare and effective models at temperature $T=0$ \cite{bulla2008numerical,PhysRevLett.99.076402}. We consider the role of finite temperature in the next subsection.
Note that the (relatively expensive) bare model `reference' observables need be computed only once. The (computationally cheap) effective model observables are iteratively calculated during the GD until Eq.~\ref{eq:aimobs} is satisfied. Once the optimized effective model parameters have been found, the effective model can be fully solved to obtain the physical properties of interest.

%%%%%%%%%%%%%%%%%%%%%%%%%%%%%%%%%%%%%%%%%
\begin{figure}[t!]
	\centering
	\includegraphics[width=\columnwidth]{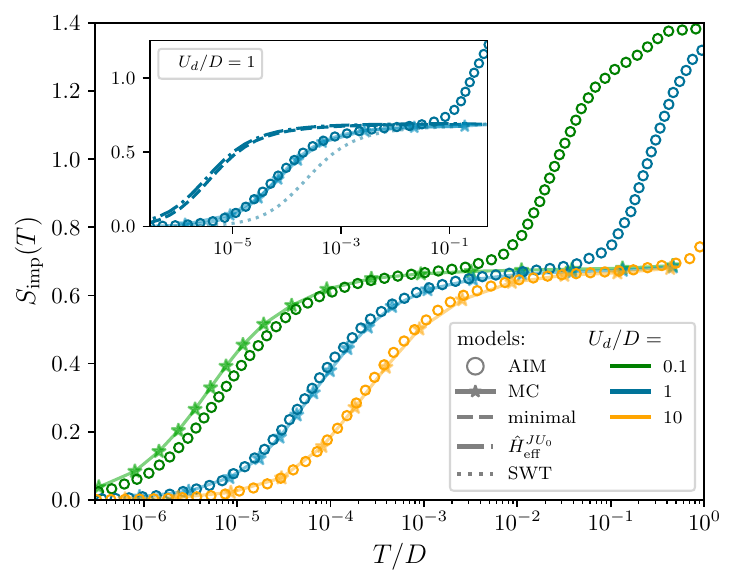}
	\caption{Demonstration of the UML optimization for the MC effective model. The impurity contribution to the entropy of the bare AIM (points) is computed as a function of $T/D$ ($D=1$) using NRG, and compared with that of the optimized MC effective model (lines) for fixed $8V^2/U_d=0.3$ and $U_d/D=0.1,1,10$. Inset shows a comparison for $U_d=1$ with the minimal Kondo model using the SWT value of $J$ (dotted) and UML-optimized $J$ (dashed), as well as UML-optimized model $\hat{H}_{\rm eff}^{JU_0}$ (dot-dashed). Training done at $T=0$ with NRG.
 }	\label{fig:aiment}
\end{figure}
%%%%%%%%%%%%%%%%%%%%%%%%%%%%%%%%%%%%%%%%%

In Fig.~\ref{fig:aiment} we show one such example calculation for the AIM to MC model mapping. The temperature dependence of the impurity entropy is shown for the bare AIM (points), compared with that of the optimized MC effective model (lines) for three different values of $U_d$, keeping $8V^2/U_d=0.3$ fixed. The inset shows a comparison with UML optimization of only $J$ in the minimal Kondo model Eq.~\ref{eq:kondo} by matching just $\langle \hat{\mathbf{S}}_d \cdot \hat{\mathbf{S}}_0 \rangle$ (dashed line), as well as UML optimization of both $J$ and $U_0$ in Eq.~\ref{eq:Himp_eff_JU} by matching $\langle \hat{\mathbf{S}}_d \cdot \hat{\mathbf{S}}_0 \rangle$ and $\langle (\hat{n}_0 -\hat{\mathbbm{1}}_{\rm eff})^2 \rangle$ (dot-dashed line). Also given for reference is the minimal Kondo model with $J$
obtained perturbatively from SWT (dotted line).

The numerical results show that UML using the MC effective model Eq.~\ref{eq:Himp_eff_P} represents a significant improvement over the other methods, already at relatively high $U_d$ (of the order of the bandwidth $U_d \sim D$) where the RG flows of the bare and effective models only mildly differ. However, we also see that the low-energy physics, characterized by the low-temperature crossover scale $T_K$, is still not perfectly reproduced by UML for $U_d/D \ll 1$. We show in Sec.~\ref{sec:ext} how the results can be systematically improved.

%%%%%%

\subsubsection{Optimizing the set of matched observables}\label{sec:optproj}
As noted above, there is some ambiguity in Eq.~\ref{eq:Hprojfull} as to which projector to eliminate in the UML observable matching process. In Eq.~\ref{eq:1emolP} we somewhat arbitrarily eliminated the projector $\hat{P}_{0,0}$ by using the identity operator, but in principle $\hat{P}_{0,1}$ or $\hat{P}_{\pm1,1/2}$ could have been eliminated. Here we present a method for identifying which projector should be eliminated, and validate this numerically by comparing effective models based on different choices. For the case of the MC effective model it turns out that indeed Eq.~\ref{eq:1emolP} is the optimal model. However for the more complicated effective models considered in Sec.~\ref{sec:ext} which feature a larger number of observables, one requires a systematic approach

%%%%%%%%%%%%%%%%%%%%%%%%%%%%%%%%%%%%%%%%%
\begin{figure}
\centering
\includegraphics[width=1.05\columnwidth]{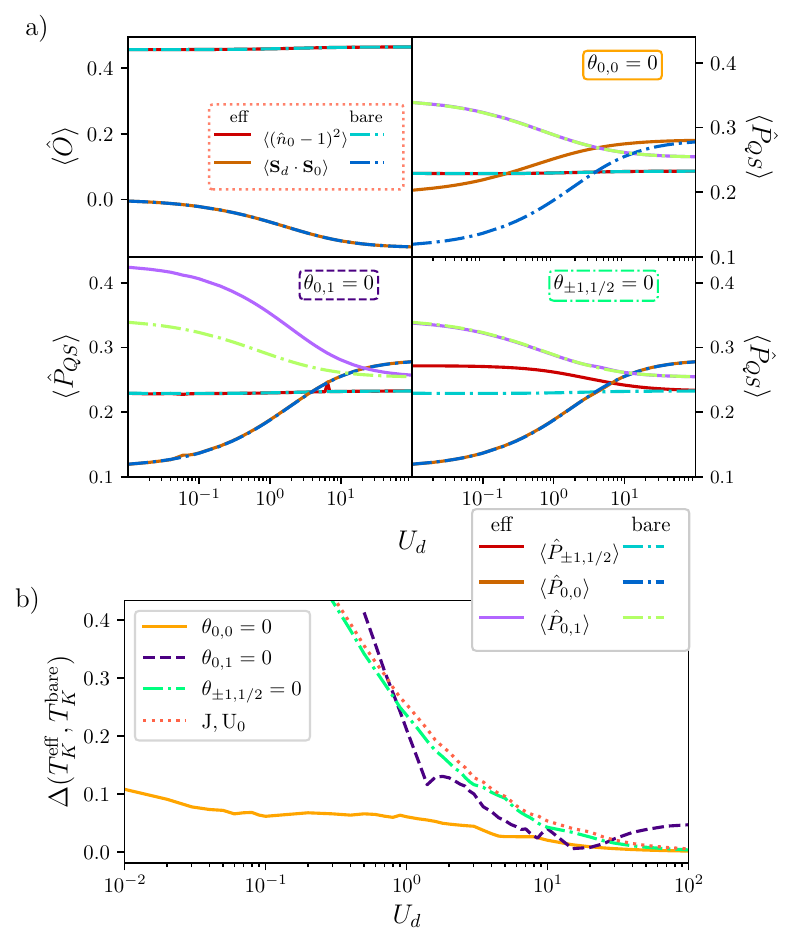}
\caption{Analysis of UML optimization of MC effective models for the AIM bare model. (a) top-left panel: $\hat{H}^{JU_0}_{\rm eff}$ (Eq.~\ref{eq:Himp_eff_JU}), with observables compared in bare and effective models, as a function of $U_d$. Other panels: similarly for $\hat{H}^{MC}_{\rm eff}$ (Eq.~\ref{eq:Hprojfull}) with $\theta_{0,0}=0$ (top-right); $\theta_{\pm1,1/2}=0$ (bottom-right); $\theta_{0,1}=0$ (bottom-left). (b) Relative (logarithmic) error for the Kondo temperature $T_K$ (Eq.~\ref{eq:error}) in the optimized model as a function of $U_d$, for the four cases considered in panel (a). The orange line corresponds to Eq.~\ref{eq:Himp_eff_P}.
Plotted for fixed $8V^2/U_d=0.3$ as a function of $U_d$, with results obtained at $T=0$ by NRG.}
	\label{fig:1eAIM}
\end{figure}
%%%%%%%%%%%%%%%%%%%%%%%%%%%%%%%%%%%%%%%%%%%%%

The first step is to define the Frobenius scalar product (FSP) on the operator space $\langle\hat{A},\hat{B}\rangle = {\rm tr}\left[\hat{A}^\dagger \cdot \hat{B}\right]$
and the corresponding norm $\| \hat{A} \| = \sqrt{\langle\hat{A}^\dagger,\hat{A}\rangle}$. \rev{Here we use the FSP to decompose an operator $\hat{h}_i$ from the effective model into contributions from the low-energy fixed point Hamiltonian of the bare model $\hat{H}^*$, and contributions from RG-irrelevant corrections to the fixed point, $\hat{\Lambda}_i$.} We define $\alpha_i=\langle\hat{h}_i,\hat{H}^*\rangle/\|\hat{H}^*\|^2$ and $\beta_i=\langle\hat{h}_i,\hat{\Lambda}_i\rangle/\|\hat{\Lambda}_i\|^2$, noting that $\langle \hat{\Lambda}_i,\hat{H}^*\rangle = 0$. The operator $\hat{\Lambda}_i$ can be found by the Gram-Schmidt procedure $\hat{\Lambda}_i = \hat{h}_i -\alpha_i\hat{H}^*$. 
We now decompose the projector as $\theta_i \hat{h}_i = \theta_i(\alpha_i \hat{H}^* +\beta_i\hat{\Lambda}) \equiv \gamma_i \hat{H}^* + \lambda_i \hat{\Lambda}_i$, which can be used for the substitution, 
\begin{align}
\label{eq:linear}
&\frac{\partial}{\partial\theta_i}\mathcal{F}_{\rm bare}(\vek{C}; \theta_i)\Big\vert_{\theta_i = 0} = \nonumber\\
&\Big[\underbrace{\beta_i\frac{\partial}{\partial \lambda_i}}_{\rm RG~irrelevant} + \underbrace{\alpha_i\frac{\partial}{\partial \gamma_i}}_{\rm RG~relevant}\Big]\mathcal{F}_{\rm bare}(\vek{C};\lambda_i,\gamma_i)\Big\vert_{\gamma_i = \lambda_i = 0}  \; .
\end{align}
The coefficients $\alpha_i$ and $\beta_i$ in Eq.~\ref{eq:linear} tell us the relative weight of RG relevant and RG irrelevant terms in the decomposition of a given operator $\hat{h}_i$. In particular, the larger $\alpha_i$ the stronger $\langle \hat{h}_i \rangle$ flows under RG. For the example of the \textit{minimal} Kondo model with $\hat{h}_i=\hat{\mathbf{S}}_d \cdot \hat{\mathbf{S}}_0$, we have trivially that $\beta_i = 0$ and the operator is entirely RG relevant, $\alpha_i=1$.  The projectors $\hat{P}_{QS}$ in \ceqn{eq:1emolP} on the other hand have a non-trivial decomposition. To render the overlap parameter $\alpha_i$ comparable for different projectors, we introduce the normalized quantity, $\tilde{\alpha}_i =|\langle\hat{h}_i,\hat{H}^*\rangle|/\|\hat{H}^*\|\| \hat{h}_i \|$. This provides a means to compare the strength of the renormalization of the effective interactions. Since we wish to compare only the most weakly renormalized observables in bare and effective models, we can use the FSP to eliminate the projector with the largest $\tilde{\alpha}_i$. For the effective interactions of $\hat{H}^{JU_0}_{\rm eff}$ and $\hat{H}^{MC}_{\rm eff}$ and the fixed point Hamiltonian $\hat{H}^* = \hat{\mathbf{S}}_d \cdot \hat{\mathbf{S}}_0$, the FSP yields,
\begin{center}
\begin{tabular}{c|c|c|c|c|c}
	\label{tab:alpha}
	& $~\hat{\mathbf{S}}_d \cdot \hat{\mathbf{S}}_0~$ &$~\left(\hat{n}_0 -1\right )^2~$  & $~\hat{P}_{0,1}~$ &  $~\hat{P}_{0,0}~$  &  $~\hat{P}_{\pm1,1/2}~$  \\
	\hline
	$\tilde{\alpha}\approx~$	& $1$  & $0$ & $0.5$ & $0.87$ & $0$  \\
\end{tabular}
\end{center}
Based on these calculated values of $\tilde{\alpha}_i$, the best effective Hamiltonian is $\hat{H}^{MC}_{\rm eff}$ with $\theta_{0,0} = 0$, since this excludes the term that gets renormalized the strongest. This is precisely Eq.~\ref{eq:1emolP}. 

We put this hypothesis to the test by performing the UML optimization for all four effective Hamiltonians and comparing the results: Eq.~\ref{eq:Himp_eff_JU} and Eq.~\ref{eq:Hprojfull} with each of the three projectors eliminated in turn. Our numerical results are presented in \cfig{fig:1eAIM}.

In the top-left of panel (a), we perform UML optimization on $\hat{H}^{JU_0}_{\rm eff}$, matching the two effective model operators $ \hat{\mathbf{S}}_d \cdot \hat{\mathbf{S}}_0 $ and $ (\hat{n}_0-1)^2$. The $T=0$ expectation values of these operators evaluated in bare and effective models using NRG are plotted as a function of $U_d$. Both observables can be matched in the effective model for all bare model parameters. In the top-right panel of \cfig{fig:1eAIM}(a) we show instead the UML optimization of $\hat{H}^{MC}_{\rm eff}$ (Eq.~\ref{eq:Hprojfull}) with $\theta_{0,0}=0$. Here we match $\langle \hat{P}_{0,1}\rangle$ and $\langle \hat{P}_{\pm1,1/2}\rangle$, which is also achievable for all $U_d$. Although $\langle \hat{P}_{0,0}\rangle$ is not matched during the optimization, we can still compare its calculated expectation value in the optimized effective model with that of the bare model. We find, as expected, that they only agree in the perturbative limit $U_d \gg 1$. It is not possible to match all projectors simultaneously in general. The bottom-left and bottom-right panels of \cfig{fig:1eAIM}(a) correspond to optimization of Eq.~\ref{eq:Hprojfull} with $\theta_{0,1}=0$ and $\theta_{\pm1,1/2}=0$, respectively -- a similar result pertains.

In \cfig{fig:1eAIM}(b) we compare the accuracy of these differently-optimized effective models, by calculating their Kondo temperature $T_K$, and comparing with that in the bare model. Our figure of merit is the (logarithmic) relative error, defined
\begin{equation}
	\label{eq:error}
	\Delta(T^{\rm eff}_K,T^{\rm bare}_K) = \left\vert 1 - \frac{\ln\big[\max(T^{\rm eff}_K,T^{\rm bare}_K)\big]}{\ln\big[\min(T^{\rm eff}_K,T^{\rm bare}_K)\big]}\right\vert \; .
\end{equation}
Given the exponential dependence of the Kondo temperature on the coupling $J$ in the minimal Kondo model \cite{hewson1993}, Eq.~\ref{eq:error} captures the error in the effective $J$. 
Our numerical results in \cfig{fig:1eAIM}(b) show that including the $U_0$ term as a correction to the Kondo model as per Eq.~\ref{eq:Himp_eff_JU} (dotted line) does not appreciably improve on the results of the minimal Kondo model. Furthermore, UML optimization of the MC effective model Eq.~\ref{eq:1emolP}, with $\theta_{0,1}=0$ (dashed) or $\theta_{\pm1,1/2}=0$ (dot-dashed), does not produce markedly better results for $\Delta(T^{\rm eff}_K,T^{\rm bare}_K)$. We attribute this to the injudicious attempt to match observables that flow strongly under RG. However, when the most RG-relevant operator is eliminated, $\hat{P}_{0,0}=0$ (orange line) the Kondo temperature is well approximated at all $U_d$. This shows an enormous improvement over the observable-matching minimal model. We conclude that the FSP is a good way to establish RG-relevance of Hamiltonian terms and to identify ill-fitted interactions for the optimization. Observables with weaker flow are better suited for the KLD distinguishability measure. This analysis provides a guide as to the proper set of observables to match within UML.

%%%%%%%%%%%%%%%%%%%%%%%%%%%%%%%%%%

\subsubsection{Temperature dependence of the optimization}

%%%%%%%%%%%%%%%%%%%%%%%%%%%%%%%%%%%%%%%%%
\begin{figure}[t]
	\centering
	\includegraphics[width=0.8\columnwidth]{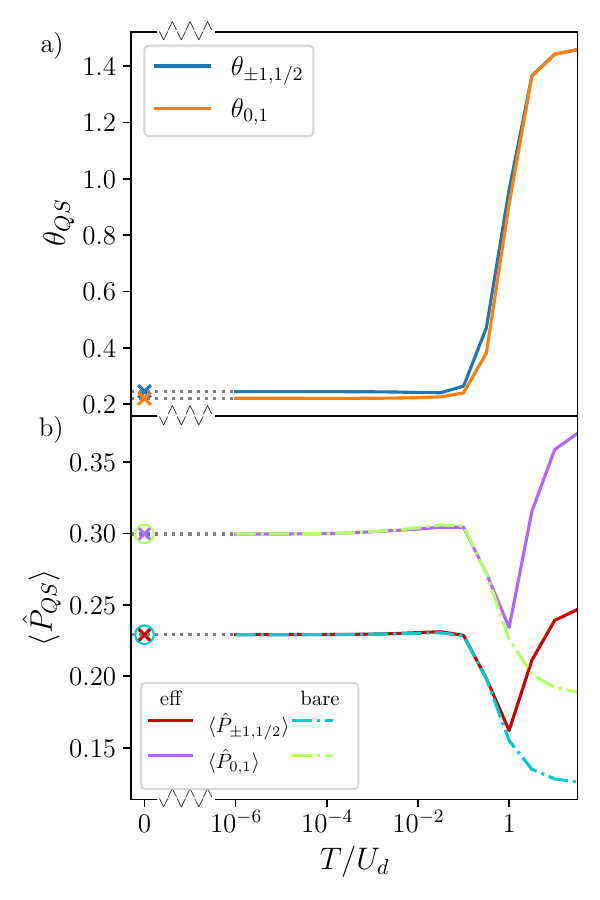}
	\caption{(a) Evolution of optimized parameters of the MC effective model Eq.~\ref{eq:Himp_eff_P} with temperature for the AIM with $U_d=0.4D$ and $8V^2/U_d=0.3$. (b) Corresponding observables evaluated in the bare and effective models. The logarithmic temperature axis includes a cut to $T = 0$ to show that optimization can be done using ground state properties if desired. }
	\label{fig:temp_evo}
\end{figure}
%%%%%%%%%%%%%%%%%%%%%%%%%%%%%%%%%%%%%%%%%

Minimization of the KLD loss function during UML involves computing static thermal expectation values, and comparing these in bare and effective models. In principle this can be done at any temperature, $T$. However, one obtains meaningful results only when the temperature is sufficiently low that the \textit{active} Fock space of the bare model is comparable to that of the effective model, $T\ll E_{\rm ex,1}-E_{GS}$. For the AIM, this means in practice $T\ll U_d$, since then impurity charge fluctuations are frozen and the description of the impurity as a local moment spin-$\tfrac{1}{2}$ degree of freedom is applicable.

So far we performed UML by calculating observables at zero temperature using NRG. We note that other methods operating at $T=0$ that target the ground state can similarly be used, for example quantum chemical methods \cite{calvo2024theoretical}, the density matrix renormalization group (DMRG) \cite{da2008transport,holzner2009kondo,zhu2024towards}, or \textit{ab initio} methods based on the variational principle \cite{foulkes2001quantum,bartlett2007coupled}. However, here we show that that the results of the effective model optimization are rather insensitive to temperature provided $T\ll U_d$. The important consequence of this is that the bare model need not be completely solved down to $T=0$, or the exact ground state of the many body system determined. Methods based on imaginary time evolution that favour relatively high temperatures $T \gtrsim 10^{-2}D$ \cite{SETH2016274} can therefore be used for complex bare model calculations. Indeed, imaginary time \textit{ab intio} methods can also be used for a first-principles treatment of the bare model \cite{sugiyama1986auxiliary,motta2018ab}.

In Fig.~\ref{fig:temp_evo} we plot the temperature-dependence of the optimized MC effective model parameters $\theta_{QS}$ (panel a) and the corresponding thermal observables $\langle \hat{P}_{QS}\rangle$ (panel b). For this demonstration, we use the AIM as the bare model and use NRG as our impurity solver. Once the AIM reaches the local moment regime for $T/U_d \lesssim 0.1$, the optimized parameters $\theta_{QS}$ become essentially temperature independent. The observables can also be precisely matched in this regime, down to $T=0$. Demonstrating that the calculated observables can be matched and do not change appreciably with temperature provides an independent check on the convergence of the UML optimization.
By contrast, it is clear that a good effective model cannot be formulated for $T/U_d \gtrsim 1$.

%%%%%%

\subsubsection{UML for complex impurities}\label{sec:complex}
Finally, we turn to UML optimization of the MC effective model for more complex impurity systems. For a bare model with a unique doublet ground state, Eq.~\ref{eq:Himp_eff_P} still constitutes a good effective model and UML involves matching the observables as per Eq.~\ref{eq:aimobs}. With conserved charge $Q$ and spin $S$ in the bare model, we can again define projectors $\hat{P}_{QS}$. However, care must now be taken because (i) a multi-orbital bare model might have a reference charge $\tilde{Q}$ that is not the same as in the MC effective model; (ii) in general the bare model will have several multiplets for a given symmetry subspace defined by the quantum numbers $Q$ and $S$; and (iii) the admissibility operator projects onto the physical impurity $S=\tfrac{1}{2}$ space of $\mathcal{H}_{\rm bare}^{\rm imp}$ but the operators are defined in the extended impurity Fock space $\tilde{\mathcal{H}}_{\rm bare}^{\rm imp}$. 

In the following we address these issues by interpreting the operators $\hat{\mathcal{P}}_{QS}=\hat{\Omega}^{\dagger}_{\rm ad}\: \hat{P}_{QS} \: \hat{\Omega}^{\phantom{\dagger}}_{\rm ad}$ in the bare model as meaning the projector onto the spin-$S$ subspace of the \textit{enlarged} impurity Fock space $\tilde{\mathcal{H}}_{\rm bare}^{\rm imp}$ with spin $S=\tfrac{1}{2}$ on the physical impurity Fock space ${\mathcal{H}}_{\rm bare}^{\rm imp}$, where the local bath orbital $\hat{c}_{0\sigma}$ has a charge $Q$ relative to half-filling. We assume a trace over different charge states and multiplets of the physical impurity space that are compatible with these conditions.
Thus, we may define,
\begin{equation}
\begin{split}
\langle\hat{\mathcal{P}}_{\pm1,1/2}\rangle_{\rm bare} &=\langle 
(\hat{n}_0-1)^2 \rangle_{\rm bare}^{S=1/2} \;, \\
\langle\hat{\mathcal{P}}_{0,1}\rangle_{\rm bare} &=\langle 
(\hat{n}_0-2\hat{n}_{0\uparrow}\hat{n}_{0\downarrow})(\tfrac{5}{4}-\tfrac{1}{3}\hat{\mathbf{S}}_{\rm imp}^2)\rangle_{\rm bare}^{S=1} \;,
\end{split}
\end{equation}
where $\hat{\mathbf{S}}_{\rm imp}=\sum_i \hat{\mathbf{S}}_i$ is an operator for the total spin of the physical impurity space and $\langle\dots\rangle_{\rm bare}^S$ denotes the thermal expectation value in the full bare model, of an operator defined in the spin $S$ block of the enlarged MC impurity Fock space. 

We emphasize that any method that can compute such observables can be used for the UML optimization. This can be done at any temperature where bare and effective models should agree (that is, $T\ll E_{\rm ex,1}-E_{GS}$ and including $T=0$ ground state calculations). For example, QMC can be used to compute reference observables in the bare model at relatively high temperatures, and NRG can be used to iteratively refine the MC effective model via GD. With the optimized effective model at hand, one can then solve the effective model down to $T=0$ cheaply with NRG, and calculate real-frequency dynamical correlation functions as might be needed for a DMFT calculation, electrical conductance via the Kubo formula, or the temperature-dependence of thermodynamic quantities such as entropy or magnetic susceptibility.

\rev{Ultimately, we envision that UML will be integrated with \textit{ab initio} methods to tackle problems that are currently out of reach for brute-force calculations. This is beyond the scope of the present work, but we provide a discussion of how few-body approximants to bare-model observables might play a role in UML in Appendix~\ref{app:abinitio}.}

%%%%%%%%%%%%%%%%%%%%%%%%%%%%%%%%%%%%%%%%%
\begin{figure}[t]
	\centering
	\includegraphics[width=0.87\columnwidth]{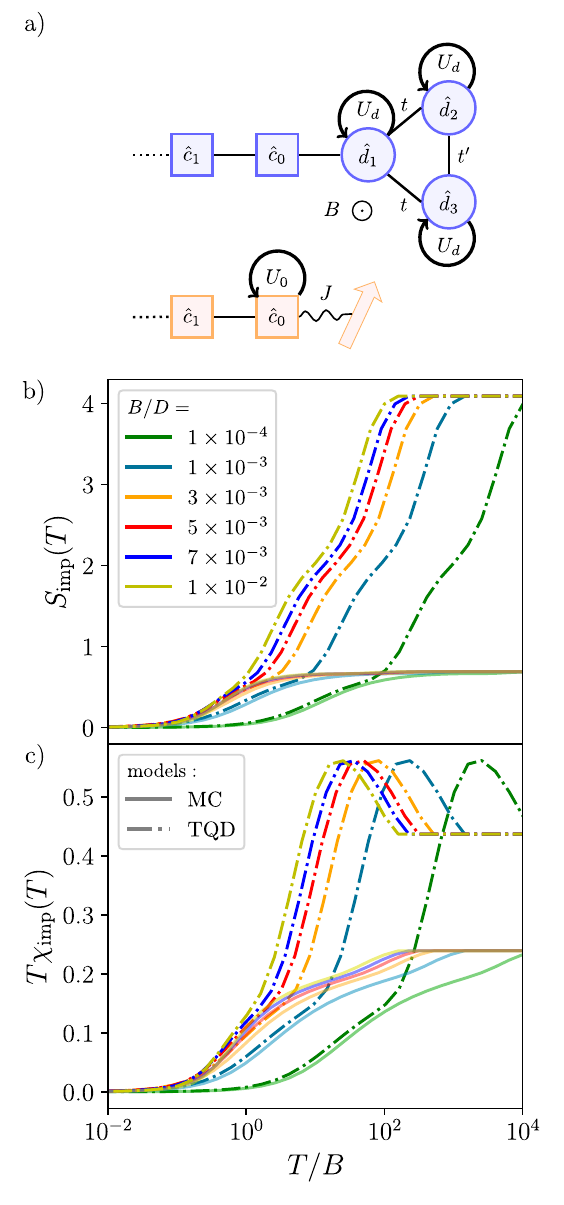}
	\caption{\rev{(a) Schematic of the triple quantum dot bare model (blue) and the associated MC effective model (orange). (b) Entropy $S_{\rm imp}(T)$ vs $T/B$ for the TQD bare model (dot-dashed lines) and the UML-optimized effective model (solid lines), for different magnetic field strengths $B$. (c) Corresponding behavior of the magnetic susceptibility $T\chi_{\rm imp}(T)$.  
 Plotted for TQD model parameters $U_d = 1.1$, 
$V = 0.25$, $t=0.10$, $t'=0.15$, $\zeta = 0.05$, $D=1$. Both bare and effective models  solved using NRG. Quantitative agreement is found in all cases in the regime $T\ll U_d$ where the mapping is expected to hold.}}
	\label{fig:tqd}
\end{figure}
%%%%%%%%%%%%%%%%%%%%%%%%%%%%%%%%%%%%%%%%%

\subsubsection{Application: triple quantum dot device}\label{sec:complex}
\rev{As a nontrivial demonstration of the UML methodology, we now consider its application to a bare model describing a triangular triple quantum dot (TQD) coupled to a metallic lead, in a magnetic field -- see Fig.~\ref{fig:tqd}(a). This system supports a rich range of physics \cite{mitchell2009quantum,*mitchell2013local,numata2009kondo,mitchell2010two}. The physical impurity Hilbert space has dimension $d_{\mathcal{H}}=4^3=64$, and SU(2)-spin symmetry as well as particle-hole symmetry is broken. We consider a regime of filling in which the TQD hosts a net spin-$\tfrac{1}{2}$ moment, delocalized across the three interacting quantum dot sites. 
The bare model Hamiltonian is given by,
\begin{align}\begin{split}
	\label{eq:tqd}
	&\hat{H}^{\text{TQD}}_{\text{bare}} = \hat{H}^{\text{bath}}_{\rm disc} +\tfrac{1}{2}U_d\sum^3_{i=1} (\hat{n}^d_i -1)^2 + \zeta \hat{N}^d + B \hat{S}^d_z \\
	&+ \sum_\sigma  \big[t \hat{d}^\dagger_{1\sigma }(\hat{d}_{2\sigma } + \hat{d}_{3\sigma }) + t' \hat{d}^\dagger_{2\sigma} \hat{d}_{3\sigma } + \hc \big] \\
	&+ V\sum_{\sigma} ( \hat{d}_{1\sigma }^{\dagger}\hat{c}_{0\sigma}^{\phantom{\dagger}}+ \hat{c}_{0\sigma}^{\dagger}\hat{d}_{1\sigma }^{\phantom{\dagger}} ) \;,
\end{split}\end{align}
where $\hat{n}_i^d=\sum_{\sigma}d_{i\sigma}^{\dagger}d_{i\sigma}^{\phantom{\dagger}}$ is the number operator for each quantum dot,  $\hat{N}^d=\sum_i \hat{n}^d_i$ is the total TQD number operator, and $\hat{S}_z^d=\tfrac{1}{2}\sum_i (d_{i\uparrow}^{\dagger}d_{i\uparrow}^{\phantom{\dagger}}-d_{i\downarrow}^{\dagger}d_{i\downarrow}^{\phantom{\dagger}})$ is the total TQD spin projection. }

\rev{The UML analysis for the TQD proceeds similarly to that of the AIM. However, with the application of a magnetic field $B$, the spin multiplets previously labelled by the quantum number $S$ are now no longer $2S+1$-fold degenerate (although total $S_z$ is still conserved). Due to the broken particle-hole symmetry, $Q=+1$ and $-1$ sectors are no longer degenerate. The lower symmetry of the bare TQD model therefore implies that we also have spin and charge anisotropies in the MC effective model. As a consequence, UML for the TQD entails matching 8 independent observables,  rather than 3 for the symmetric AIM. 
Thermal expectation values of the symmetry projectors $\hat{P}_{Q,S_z}$ are the observables which we compare in bare and effective models, with $(Q,S_z)=(0,0)$, $(0,1)$, $(0,-1)$, $(1,1/2)$, $(1,-1/2)$, $(-1,1/2)$, $(-1,-1/2)$. The $(0,0)$ block is a diagonal $2\times 2$ matrix. These observables can be defined and computed (here using NRG) in both bare and effective models.}

\rev{The effective model is then optimized by minimizing the KLD loss function by gradient descent. Since the KLD is still convex and its gradient, Eq.~\ref{eq:gd}, is expressed in terms of the computed observables $\langle \hat{P}_{Q,S_z}\rangle$, the optimization remains highly efficient, despite the increase from 3 to 8 in the number of parameters to be tuned. The gradient descent method is easily able to treat optimization problems of this complexity \cite{bengio}.  Indeed, as shown later, systems with a far larger number of parameters in the non-convex case are also readily optimized. }

\rev{Each step of the optimization process involves solving the MC effective model using the updated coupling constants, and computing the updated set of observables. The compute time simply grows linearly in the number of observables. These are compared with those of the bare model from a one-shot calculation. Even though the effective model has to be solved many times, these calculations are comparatively cheap, since by construction the  Hilbert space dimension of the impurity in the effective model is less than that in the bare model. Although the effective model may have a larger number of coupling constants than the original bare model, it is defined in a reduced Hilbert space and therefore is of lower computational complexity to solve.}

\rev{The iterative solution and refinement of the effective model with UML has essentially fixed computational cost, and is highly efficient. The potential bottleneck is in the one-shot solution of more complex bare models, as noted above. For the TQD however, NRG can still be used to obtain exact benchmark results.}

\rev{We demonstrate the viability of UML for the TQD, Eq.~\ref{eq:tqd}, in Fig.~\ref{fig:tqd}. Thermodynamic properties of the bare TQD (dot-dashed lines) and the optimized MC effective model (solid lines) are compared, for different temperatures $T$ and field strengths $B$. Panel (b) shows the impurity contribution to the total entropy, $S_{\rm imp}(T)$, and (c) shows the magnetic susceptibility $T\chi_{\rm imp}(T)$, both vs $T/B$. The simplified effective model successfully captures the physics of the TQD for $T\ll U_d$. At higher temperatures, TQD degrees of freedom not included in the effective model contribute to thermodynamic averages. }

\rev{As $B$ is increased in the bare TQD model, we see a crossover from Kondo-screening of the TQD local moment, to polarization of the local moment by the field. Both the entropy and the magnetic susceptibility are fully quenched as $T\to 0$ for any $B$. However, the scale on which this happens depends on the underlying mechanism. For small fields $B\ll T_K$ (green line) screening occurs due to the Kondo effect, and the physics is insensitive to $B$. However, at larger $B$ the Kondo effect, which depends on dynamical spin-flip scattering, is destroyed. The TQD local moment is then screened on the scale of $B$. For $B\gg T_K$ we therefore see scaling collapse in terms of $T/B$, with polarization of the local moment setting in for $T/B\sim 1$ (see yellow, blue, red lines). The low-energy behavior of the optimized effective model captures this crossover and the onset of scaling vs $T/B$ accurately. }

\rev{We conclude that more complex models, including those with more interacting degrees of freedom and lower symmetries, can be treated efficiently and accurately with UML. Physical properties of the resulting optimized effective models agree quantitatively at low temperatures with bare model benchmark reference results.}

%%%%%%%%%%%%%%%%%%
%%%%%%%%%%%%%%%%%%

\section{Systematic improvement of extended models}\label{sec:ext}

We now put the above discussions on a more general footing, and consider systematic improvements to the MC effective models explored so far.

The expressibility of our effective model to capture different aspects of the physics of the bare model depends on the types of operators that can be included. Going beyond the minimal Kondo model, which includes only the most RG-relevant operator, our MC effective model lives in an extended impurity Fock space Eq.~\ref{eq:newfockeff}, which allows for inclusion of more general coupling terms, Eq.~\ref{eq:Himp_eff_P}. Our starting point in this section is to note that the quality of the effective model can be systematically improved by further enlarging the effective impurity Fock space:
\begin{equation}\label{eq:Fockext}
\dbtilde{\mathcal{H}}^{\rm imp}_{\rm eff} =
\mathcal{H}^{\rm imp}_{\rm eff}  \otimes \mathcal{H}^{\rm bath}_{\rm 0} \otimes \mathcal{H}^{\rm bath}_{\rm 1}  \;,
\end{equation}
which now includes both $\hat{c}_{0\sigma}$ and $\hat{c}_{1\sigma}$ bath orbitals. The extended effective model then takes the form,
\begin{equation}\label{eq:Hext}
\hat{H}^{\rm ext}_{\rm eff}(\pmb{\theta}) = \hat{\dbtilde{H}}^{\rm imp}_{\rm eff}(\pmb{\theta})+\hat{H}^{\rm bath}_2+ \hat{H}^{\rm hyb}_2 \;,
\end{equation}
where the extended effective impurity Hamiltonian $\hat{\dbtilde{H}}^{\rm imp}_{\rm eff}(\pmb{\theta})$ lives in the extended impurity Fock space $\dbtilde{\mathcal{H}}^{\rm imp}_{\rm eff}$ and correspondingly the bath and hybridization terms have been redefined according to Eqs.~\ref{eq:newbath}, \ref{eq:newhyb}. The bath chain now starts on site index $N=2$. The impurity part can of course be further extended if required. The MC effective model discussed in Sec.~\ref{sec:minconst} can be viewed as the first member of a family of such extended models.

When the original bath Eq.~\ref{eq:bath} takes the form of a Wilson chain \cite{KWW}, we note that the $\hat{c}_{n\sigma}$ operators become progressively less RG relevant as the chain index $n$ increases (specifically, the most relevant operator of the linearized RG transformation \ceqn{eq:the_linearlised_RG} involved in the decomposition of $\hat{c}_{n\sigma}$ near the low-energy fixed point has an eigenvalue $|\lambda^*|$ that decreases with $n$). Therefore, expanding our effective impurity Hamiltonian to include operators from Wilson orbitals with larger $n$ results in additional terms that are progressively more RG irrelevant. On the one hand, strongly RG irrelevant corrections to our effective model become less important for determining the low-energy physics, but on the other hand, expectation values of these operators flow very weakly under RG and therefore can be accurately matched even at relatively high energy scales with corresponding observables in the bare model during UML optimization.

To proceed with the construction and UML optimization of our generalized effective models, we first note that the
symmetry representations of U(1) charge ${\rm Q}$ and SU(2) spin ${\rm S}$ on $\dbtilde{\mathcal{H}}^{\rm imp}_{\rm eff}$ still commute, and so we can construct a basis on the effective Fock space labelled by quantum numbers $S$ for the non-abelian spin, $S_z$ for the abelian magnetic quantum number of the Cartan subalgebra of ${\rm S}$, and $Q$ for the abelian charge, 
\begin{align}
	\bra{Q,S,S_z;m}{Q',S',S_z';m'}\rangle = \delta_{Q Q'}\delta_{S S'}\delta_{S_z S_z'}\delta_{m m'} \; ,
\end{align}
where $m$ is the \textit{multiplet} index that distinguishes different basis states with the same set of quantum numbers $Q,S,S_z$. Unlike in the basic MC example, in general the symmetry blocks of the extended impurity Hamiltonian are not simply $1\times 1$ scalars.
Furthermore, such multiplets are not unique since linear combinations of different multiplets with the same set of quantum numbers are also valid multiplets. This introduces a gauge degree of freedom in the choice of the basis that means we cannot unambiguously compare degenerate multiplets between bare and effective models based on symmetry alone. We show below that fixing the gauge by lifting the multiplet degeneracy allows to reestablish a one-to-one correspondence between bare and effective models.

Accounting for the multiplet structure of the symmetry spaces, we construct projectors by generalizing Eq.~\ref{eq:Pqs},
\begin{align}
	\hat{P}_{QS} = \sum_{S_z,m} \ket{Q,S,S_z;m}\bra{Q,S,S_z;m} \;.
\end{align}
One can use these projectors as a mutually commuting operator basis to construct the effective impurity Hamiltonian,
\begin{align}
	\label{eq:eml_proj_ham}
\hat{\dbtilde{H}}^{\rm imp}_{\rm eff}(\pmb{\theta}) = \sum_{Q,S} \theta_{QS} \hat{P}_{QS} \; .
\end{align}
The allowed quantum number configurations compatible with Eq.~\ref{eq:Fockext} are,
\begin{center}
\begin{tabular}{ |c|c|c|c| } 
 \hline
 $~~Q~~$ & $~~S~~$ & $~~~S_z~~~$ & $~~M_{\rm eff}~~$ \\ 
 \hline \hline
$0$ & $\tfrac{1}{2} $ & $\pm \tfrac{1}{2} $ & $4$ \\
 \hline
$0$ & $\tfrac{3}{2} $ & $~\pm\tfrac{1}{2},~\pm\tfrac{3}{2}~$ & $1$ \\
 \hline
 $\pm 1$ & $0$ & $0$ & $2$ \\
 \hline
 $\pm 1$ & $1$ & $0,~\pm1$ & $2$ \\
 \hline 
 $\pm 2$ & $\tfrac{1}{2}$ & $\pm\tfrac{1}{2}$ & $1$ \\
\hline
\end{tabular}
\end{center}
where the charge $Q$ is again specified relative to half-filling, $M_{\rm eff}$ is the number of multiplets with a given $Q$ and $S$, and we take the impurity to be a spin-$\tfrac{1}{2}$ object. We therefore have 15 $QS$ multiplets that span the extended impurity space.

The advantage of formulating the effective model in terms of such symmetry projectors is that equivalent projectors can be identified in the bare model, since the two models share the same symmetries. The projectors so defined also mutually commute, which means the UML optimization problem is convex (Appendix~\ref{app:convexity}).

A disadvantage is that several $QS$ blocks contain more than one multiplet, which means that the projectors cannot distinguish between multiplets with the same $QS$. The effective model Eq.~\ref{eq:eml_proj_ham} is therefore \textit{not} the most general extended impurity model one can write down. Interactions in the bare model will in general lift the energy degeneracy of these multiplets, but this effect cannot be captured by a model of the form of Eq.~\ref{eq:eml_proj_ham}.

%%%%%%%

\subsection{Fixed basis}
A partial solution is to fix the basis according to physical intuition of the problem, and add terms to Eq.~\ref{eq:eml_proj_ham} that lift the energy degeneracy of multiplets with the same $Q$ and $S$ quantum numbers. This should be done in such a way that: (i) the new set of operators still all mutually commute, so as to guarantee convexity of the UML optimization problem; and (ii) the added terms have a physical correspondence in the bare model. This approach is still not the most general form of extended MC effective model, because we make a basis choice. We return to the fully general case later.

Since the $\hat{c}_{0\sigma}$ and $\hat{c}_{1\sigma}$ bath orbitals are common to both bare and effective models, and there is a physical tunneling matrix element between these orbitals in both cases, we introduce the operator,
\begin{align}
	\hat{T} = \sum_\sigma  (\hat{c}^\dagger_{0\sigma }\hat{c}_{1\sigma }^{\phantom{\dagger}} + \hat{c}^\dagger_{1\sigma }\hat{c}_{0\sigma }^{\phantom{\dagger}} ) \; .
\end{align}
The tunneling operator $\hat{T}$ has support only on the physical bath Fock space and therefore does not depend on the structure of the impurity in the bare model. However, one might anticipate that $\langle \hat{T}\rangle_{\rm bare}$ will depend on details of the bare impurity model. Action of the tunneling operator conserves charge and spin and so $\hat{T}$ commutes with $\hat{Q}$ and $\hat{\mathbf{S}}^2$ defined on the extended impurity space. We can use the eigenvalues $T$ of the tunneling operator to label our basis states, as shown explicitly in Appendix~\ref{app:double}. With basis states $|Q,S,T,S_z;m\rangle$ we almost fully remove multiplet degeneracies,
\begin{center}
\begin{tabular}{ |c|c|c|c|c| } 
 \hline
 $~~Q~~$ & $~~S~~$ & $~~T~~$ & $~~~S_z~~~$ & $~~M_{\rm eff}~~$ \\ 
 \hline \hline
$0$ & $\tfrac{1}{2}$ & 0 & $\pm\tfrac{1}{2}$ & $2$ \\
 \hline
$0$ & $\tfrac{1}{2}$ & $\pm 2$ & $\pm\tfrac{1}{2}$ & $1$ \\
 \hline
$0$ & $\tfrac{3}{2}$ & 0 & $~\pm\tfrac{1}{2},~\pm\tfrac{3}{2}~$ & $1$ \\
 \hline
 $\pm 1$ & $0$ & $\pm1$ & $0$ & $1$ \\
 \hline
 $\pm 1$ & $1$ & $\pm1$ & $0,~\pm1$ & $1$ \\
 \hline 
 $\pm 2$ & $\tfrac{1}{2}$ & 0 & $\pm\tfrac{1}{2}$ & $1$ \\
\hline
\end{tabular}
\end{center}
The $(Q,S,T)=(0,\tfrac{1}{2},0)$ subspace still has $M_{\rm eff}=2$ multiplets, but this degeneracy can be lifted by introducing an additional operator (which we take to act only in this subspace), $\hat{W}=\hat{n}_{0\uparrow}\hat{n}_{0\downarrow}+\hat{n}_{1\uparrow}\hat{n}_{1\downarrow}$, which measures double-occupancy on the bath sites included in the extended impurity space. $\hat{W}$ has eigenvalues 0 or 1 only in the subspace where the operator acts, so we employ a concise notation in which the $(T,W)$ eigenvalues are augmented into one label $(T,W) \mapsto T$.
As such, all multiplet degeneracies are lifted when we express
\begin{align}
	\label{eq:HextQST}
	\hat{\dbtilde{H}}^{\rm imp}_{\rm eff}(\pmb{\theta}) = \sum_{Q,S,T} \theta_{QST} \hat{P}_{QST} \; ,
\end{align}
where the spin multiplets $|Q,S,T,S_z\rangle$ are now unique (we drop the multiplet index $m$) and $\hat{P}_{QST}=\sum_{S_z}|Q,S,T,S_z\rangle\langle Q,S,T,S_z|$.

Fixing the multiplet basis in such a way allows to address each multiplet individually and make a comparison of their projector expectation values in bare and effective models. This is the best one can do while retaining a fully commuting set of extended impurity Hamiltonian operators, as required for convexity. However, the most general extended effective model would not constrain the basis in this way. 

\rev{If desired, the projectors $\hat{P}_{QST}$ can be formulated in second-quantized form (see Appendix~\ref{app:abinitio}). New types of coupling term arise in the extended MC effective model, relative to the basic MC model. In particular, we see \textit{assisted hopping} terms, in which an impurity spin-flip is accompanied by electronic tunneling between bath sites. This might be expected from the physical interpretation of the Kondo effect in terms of spin-flip scattering.}

%%%%%%%%%%

\subsection{Generalized basis}
The most general extended effective impurity model must include off-diagonal terms coupling different multiplets. This can be captured by operators of the type
\begin{equation}\label{eq:Xop}
\hat{X}_{QS}^{T,T'} = \sum_{S_z}|Q,S,T,S_z\rangle\langle Q,S,T',S_z|+{\rm H.c.}
\end{equation}
such that our effective model reads,
\begin{align}
	\label{eq:HextQSTTp}
	\hat{\dbtilde{H}}^{\rm imp}_{\rm eff}(\pmb{\theta}) = \sum_{Q,S}\:\sum_{T,T'} \theta_{QS}^{T,T'} \hat{X}_{QS}^{T,T'} \;.
\end{align}
The model thus features 25 distinct interaction terms. 
Due to the cross terms $T\ne T'$, the operators $\hat{X}$ do not in general commute, and so the UML optimization problem is no longer convex. This is unavoidable for more complex effective models, but the appearance of local minima in the search landscape can still be dealt with in practice using more sophisticated GD-based routines. 
%The minimization of \ceqn{eq:P_KLD} can still be carried out using GD, however local minima can be encountered. We have found that running in to local minima can be avoided by initializing the optimization with the expectation values of the bare model $ \langle \hat{P}_{QS;t,t'} \rangle_{\rm bare} \xrightarrow{\rm initialize}\theta_{QS;t,t'} $. This has the advantage that symmetries of the model are automatically respected and small bare expectation values lead directly to small effective parameters.

The impact of including the off-diagonal terms $T\ne T'$ can be quantified using the FSP. An optimized representation of any operator $\hat{Y}$ using the set of operators contained in the model Eq.~\ref{eq:HextQSTTp} can be found by minimizing the normalized FSP \cite{che2021learning},
\begin{align}
	\label{eq:forbenius}
	\mathcal{L}( \pmb{\theta} ,\hat{Y}) = 1 - \frac{\left|\Big\langle \hat{\dbtilde{H}}^{\rm imp}_{\rm eff}(\pmb{\theta}) ,\hat{Y}\Big\rangle \right|}{\Big\| \hat{\dbtilde{H}}^{\rm imp}_{\rm eff}(\pmb{\theta}) \Big\|\cdot\Big\|\hat{Y}\Big\|} \; .
\end{align}
As a demonstration of the expressibility of the model  Eq.~\ref{eq:HextQSTTp}, we consider its ability to represent the Kondo Hamiltonian operator $\hat{Y} = \hat{\mathbf{S}}_d \cdot \hat{\mathbf{S}}_0$ from \ceqn{eq:kondo}. In the limit $U_d\gg 1$ where the perturbative SWT holds, the effective model should reduce to \ceqn{eq:kondo}, so this is a nontrivial and stringent check that a good effective model should satisfy. By minimizing $\mathcal{L}( \pmb{\theta} ,\hat{\mathbf{S}}_d \cdot \hat{\mathbf{S}}_0)$ with respect to 
$\pmb{\theta}$ using standard gradient descent, we find $\mathcal{L}( \pmb{\theta}_{\rm opt} ,\hat{\mathbf{S}}_d \cdot \hat{\mathbf{S}}_0)=0$, meaning that Eq.~\ref{eq:HextQSTTp} can perfectly represent this operator. However, if we use a fixed basis (no off-diagonal terms) as with Eq.~\ref{eq:HextQST}, then $\mathcal{L}( \pmb{\theta}_{\rm opt} ,\hat{\mathbf{S}}_d \cdot \hat{\mathbf{S}}_0)\approx 0.3$. The fixed basis therefore substantially reduces the expressibility of the extended model. Therefore, even though the fully general extended model lacks convexity, we argue that this feature should be sacrificed and that Eq.~\ref{eq:HextQSTTp} should be used.

%%%%%%%%%%%%%%%%%%%

\subsection{UML optimization of extended models}

%%%%%%%%%%%%%%%%%%%%%%%%%%%%%%%%%%%%%%%%%
\begin{figure}
	\centering
\includegraphics[width=\columnwidth]{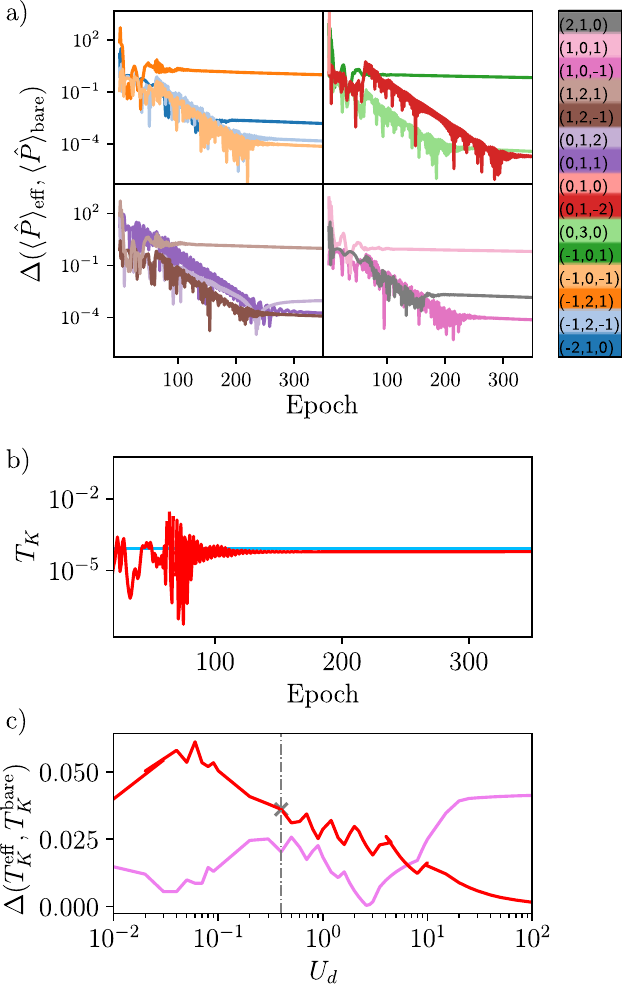}
\caption{Analysis of UML optimization of the extended MC effective models (Eqs.~\ref{eq:HextQST} and \ref{eq:HextQSTTp}), taking the AIM as our bare model. (a) Comparison of relative errors in computed observables (Eq.~\ref{eq:obserror}) as a function of the GD optimization parameter update step (epoch). Lines are color-coded according to the key, labelled by the (diagonal) projector quantum numbers $(Q,S,T)$. 
(b) Convergence of the Kondo temperature $T_K$ with epoch (red line) compared with the reference result in the bare model (blue line).
(c) Comparison of relative (logarithmic) error in the Kondo temperature for the fixed-basis effective model  Eq.~\ref{eq:HextQST} (pink line) and generalized basis model Eq.~\ref{eq:HextQSTTp} (red line).
Plotted for $8V^2/U_d=0.3$ and $U_d=0.4$, with the optimization performed at $T=0$ using NRG as the impurity solver for both bare and effective models.}
	\label{fig:training}
\end{figure}
%%%%%%%%%%%%%%%%%%%%%%%%%%%%%%%%%%%%%%%%%

Extended MC effective models with a general multiplet basis such as Eq.~\ref{eq:HextQSTTp} can represent any Hamiltonian term in the extended Fock space of Eq.~\ref{eq:Fockext}, consistent with SU(2) spin and U(1) charge symmetries.

Within UML, the GD optimization of the effective model is achieved by matching observables in the bare and effective models, Eq.~\ref{eq:gd}. As noted in Sec.~\ref{sec:minconst} however, one cannot hope to match the \textit{full} set of symmetry projectors because the optimization problem is overdetermined. The solution discussed in Sec.~\ref{sec:optproj} is to eliminate the observable that flows strongest under RG, as quantified by the FSP Eq.~\ref{eq:forbenius}.

However, even by omitting the observable with the largest $\tilde{\alpha}$, we are still left with 24 observables to compute in the effective model Eq.~\ref{eq:HextQSTTp}. It is not obvious that for generalized effective models one can always simultaneously match all such observables in bare and effective models. Indeed, we find that even in the optimized effective model, a small number of observables are not matched, and the KLD loss function can therefore only be approximately minimized. To quantify this, we introduce a measure of the relative error in a given observable,
\begin{equation}\label{eq:obserror}
	\Delta(\langle \hat{h}_i \rangle_{\rm eff},\langle \hat{h}_i \rangle_{\rm bare})  = \left\vert 1 - \frac{\max(\langle \hat{h}_i \rangle_{\rm eff},\langle \hat{h}_i \rangle_{\rm bare})}{\min(\langle \hat{h}_i \rangle_{\rm eff},\langle \hat{h}_i \rangle_{\rm bare})}\right\vert \;.
\end{equation}
When learning a representation for complex bare models (for example \textit{ab initio} models of a realistic molecular junction), the Kondo temperature $T_K$ will not typically be a good figure of merit to assess the quality of the effective model. This is because, for any practical application of UML, the bare model will not itself be solvable down to very low but finite temperatures -- hence the need in the first place for a simplified effective model that can be fully solved. Therefore one will not in general have access to the reference $T_K$ for the bare model, from which to compute the relative error Eq.~\ref{eq:error}. The error in computable observables is thus more practical; the KLD loss function gradient depends on these observables, which are matched during the optimization process. For simple bare model test cases such as the AIM, we can however still compare with $T_K$ and other quantities.

Taking the AIM again as our bare model for demonstration, we perform UML optimization on the extended MC effective model Eq.~\ref{eq:HextQSTTp}. For representative AIM parameters, in Fig.~\ref{fig:training}(a) we plot the relative error of selected (diagonal) observables labelled by the projector quantum numbers $(Q,S,T)$, as a function of the optimization step (epoch).  We find that most effective model observables converge exponentially quickly towards their reference bare model counterparts, whereas a few are essentially stationary and cannot be matched. 
\rev{Indeed, it is in general not possible to perfectly match multiple observables simultaneously, and the minimum relative error is small but remains finite after convergence. This is because the effective model has good but not unlimited expressibility: minimization of the KLD loss function typically involves a compromise in the optimized solution.}
However, as Fig.~\ref{fig:training}(b) shows, this does not make much difference to the quality of the optimized effective model, as quantified by the Kondo temperature $T_K$. After $\sim 100$ or so training steps, $T_K$ for the effective model (red line) converges quite accurately to that of the bare model (blue line).

In Fig.~\ref{fig:training}(c) we show the relative (logarithmic) error in $T_K$ in the fully  effective model, as a function of AIM interaction strength $U_d$, with the red line for the fully general extended MC effective model Eq.~\ref{eq:HextQSTTp}, compared with the fixed-basis model Eq.~\ref{eq:HextQST} as the purple line. We see that $T_K$ is reproduced rather well by the effective model in both cases, for all values of $U_d$. As expected from the above arguments, the generalized model does better at large $U_d$. However, somewhat unexpectedly, the fixed-basis model has a smaller error at small $U_d$. We attribute this to the fact that optimization of the fixed-basis model requires to match fewer observables, which is easier to achieve in practice. Furthermore, the additional (off-diagonal) observables in Eq.~\ref{eq:HextQSTTp} appear to flow more strongly under RG at small $U_d$ than the projectors in Eq.~\ref{eq:HextQST}, meaning that matching them may ironically \textit{reduce} the quality of the optimized effective model. Despite these subtleties of the training process, the extended MC effective models do represent a systematic improvement over simpler effective models.

We compare the different approaches in the following.

%%%%%%%%%%%%%%%%%%%%%%%%%%%%%%%%%
%%%%%%%%%%%%%%%%%%%%%%%%%%%%%%%%%
%%%%%%%%%%%%%%%%%%%%%%%%%%%%%%%%%

\section{Comparison of models} 
\label{sec:gml_quant_learn}

In this section we consider the relative performance of the various effective models introduced in this paper. \rev{Each is optimized by UML to capture the physics of the AIM, which serves as our benchmark bare model.} As we have shown, it is in general not possible to simultaneously capture within a simplified effective model the universal low-temperature thermodynamics (as characterized by the Kondo temperature $T_K$) as well as the behavior of static thermal expectation values of local operators. Any effective model must therefore strike a compromise in reproducing these features. Different methods may target different aspects of the physics and be better suited to different applications. 

The $\mathcal{F}$-learning approach \cite{rigo2020machine} requires calculation of the full free energy of bare and effective models within the optimization process. It guarantees to reproduce the low-energy thermodynamics and $T_K$ within a minimal Kondo model, but fails to capture local observables except in the perturbative limit $U_d\gg D$ where the SWT also holds as a good description. However, it might not be possible to accurately calculate the free energy of a complex bare model in practice, so other methods might be preferred.

UML, by contrast, is based on matching certain local observables, and so by construction these observables agree in bare and optimized effective models. However, applied to crude effective models, such as minimal Kondo-type models that only feature the most RG-relevant terms, UML yields solutions that do not capture well the low-energy thermodynamics and Kondo scale $T_K$. By including minimally constrained models in an expanded effective impurity Fock space, we systematically improve upon results obtained by UML. Such models include RG-marginal and RG-irrelevant corrections, and UML optimization is done using the set of observables that do not flow strongly.  Depending on the level of theory employed, one can optimize with UML generalized effective models that do capture local observables while also approximating $T_K$ with good accuracy.

\begin{figure}[t!]
\centering
\includegraphics[width=0.75\columnwidth]{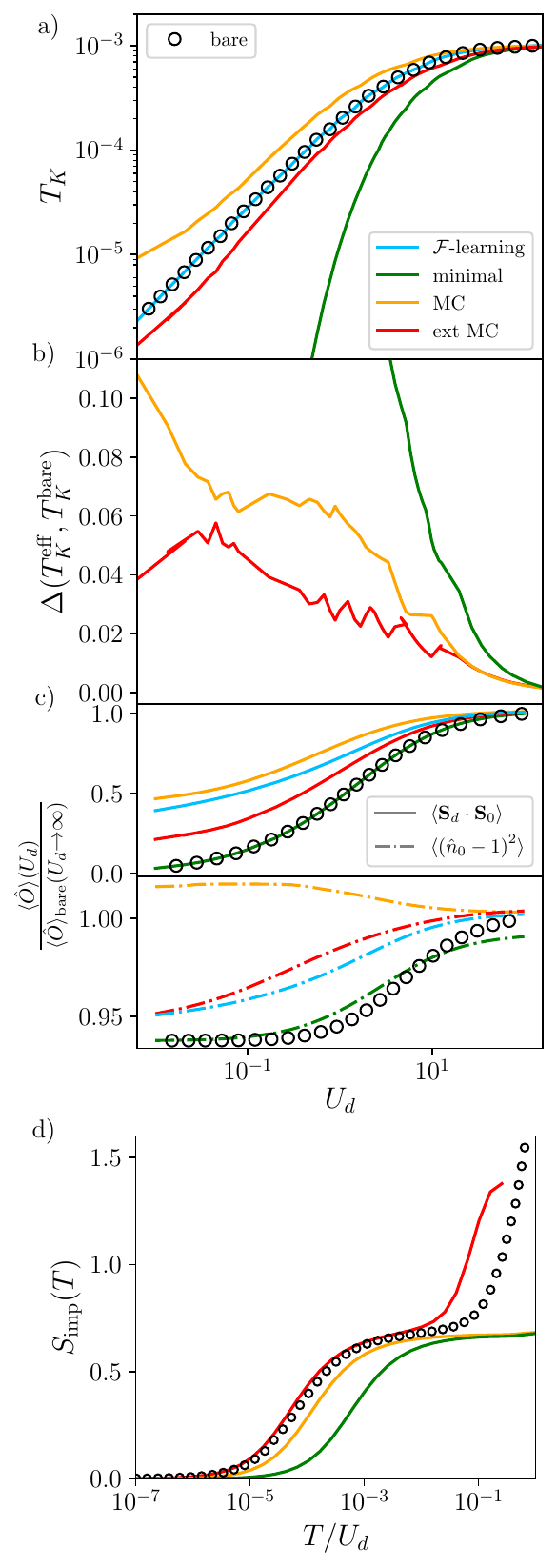}
\caption{Comparison of performance metrics for the optimized effective models considered in this work, taking the AIM as the bare model.  (a) Kondo temperature $T_K$ vs $U_d$, comparing the AIM (points) and $\mathcal{F}$-learned Kondo result (blue line) with the UML-optimized effective models: minimal (green), MC  (orange), and extended MC (red). Corresponding $T_K$ relative error shown in (b). \rev{Observables $\langle\hat{O}\rangle$ normalized by their $U_d\to\infty$ values  shown in panel (c), with $\hat{O}=\mathbf{S}_d\cdot \mathbf{S}_0$  (solid lines, top) and $(\hat{n}_0-1)^2$ (dashed lines, bottom).} (d) Entropy $S_{\rm imp}(T)$ vs $T/U_d$ for the bare AIM ($U_d=0.4$, points) and the optimized effective models (lines). We set $8V^2/U_d=0.3$ throughout.}
	\label{fig:comparison}
\end{figure}

These features are illustrated in Fig.~\ref{fig:comparison}, where we compare the performance of the different effective models.

%%%%%%%%%%%%%%%%%%%%%%%%%%%%%%%%%%%%%%%%%
%%%%%%%%%%%%%%%%%%%%%%%%%%%%%%%%%%%%%%%%%

The UML method is based on minimizing the KLD loss function \ceqn{eq:P_KLD} that quantifies the distinguishability between bare and effective model. In general ML applications, the value of the loss function is a useful metric to judge the training progress and ultimately the performance of the model. In the case of \ceqn{eq:P_KLD} however, the KLD is essentially impossible to compute in practice. The training progress cannot therefore be established in the standard way. On the other hand, we have shown that to minimize the KLD via GD we only need to know its gradient, \ceqn{eq:gd}. This can be done without actually ever evaluating the KLD itself. For a commuting set of local impurity operators (for example the symmetry projectors in MC models), the KLD is convex and so the gradient vanishes only at saddle-points or global minima of the KLD. Thus, the gradient is itself a good replacement for the loss function to determine whether the training of a model has converged. For the cases considered in Fig.~\ref{fig:comparison}, the KLD gradient can always be made to vanish. 

We therefore turn to our other metrics to quantify the relative performance of the models.  With our NRG solution of the bare and optimized effective models, we can compute the Kondo temperature directly. This is shown in Fig.~\ref{fig:comparison}(a) as a function of $U_d$. The exact result for the bare AIM (points) are compared with the effective models (lines): the $\mathcal{F}$-learning and UML result for the minimal Kondo model shown as blue and green lines; the MC and extended MC models shown as the orange and red lines. Panel (b) shows the corresponding relative error $\Delta(T^{\rm eff}_K,T^{\rm bare}_K)$ evaluated using Eq.~\ref{eq:error}. 
In panel (c) we plot the $T=0$ value of static observables $\langle \hat{O}\rangle=\langle\hat{\mathbf{S}}_d \cdot \hat{\mathbf{S}}_0\rangle$ and $\langle (\hat{n}_0-1)^2\rangle$, comparing results in bare AIM (points) with those of the optimized effective models (lines). We have normalized the results by their value in the $U_d/D\to \infty$ limit.

First, we note that by construction, $\mathcal{F}$-learning perfectly reproduces the $T_K$ value for all $U_d$, blue line in Fig.~\ref{fig:comparison}(a). However the behavior of $\langle \hat{O}\rangle=\langle\hat{\mathbf{S}}_d \cdot \hat{\mathbf{S}}_0\rangle$, which is not directly optimized by $\mathcal{F}$-learning, is poorly captured, see panel (c). On the other hand  UML for the minimal Kondo model (green line) directly matches the observable $\langle \hat{O}\rangle=\langle\hat{\mathbf{S}}_d \cdot \hat{\mathbf{S}}_0\rangle$ in panel (c) and therefore by construction agrees perfectly with the bare AIM results. The $T_K$ in this case is not directly matched, and panel (a) shows that UML for the minimal model performs very poorly using that metric, except in the SWT (perturbative) limit $U_d/D\to \infty$. 

The behavior of the Kondo temperature and the observables is a stringent nontrivial performance check for the MC and extended MC models, since neither $T_K$ nor $\langle\hat{\mathbf{S}}_d \cdot \hat{\mathbf{S}}_0\rangle$ are directly matched within UML. We find that UML represents a good compromise on the performance of these models for the two metrics. Good performance is seen for each over all $U_d$, with a systematic improvement for extended MC over plain MC.

With the optimized effective models at hand, we compute using NRG the full temperature-dependence of the thermodynamic entropy for a representative example of the bare AIM with $U_d=0.4D$, comparing the results in Fig.~\ref{fig:comparison}(d). Within the regime of validity of the effective models ($T/U_d\ll 1$), we see that the MC and extended MC models do accurately capture the physics of interest for the AIM.

%########################
%########################

\section{Conclusion} 

In this paper, we focused on developing new ML-inspired methodologies for the automatic derivation of expressive and accurate effective models for quantum impurity systems. To establish the feasibility and performance of our UML method, we studied in detail the AIM as the bare model. This is because the AIM is a very well-known paradigm \cite{hewson1993}, which we can solve exactly down to $T=0$ using NRG \cite{bulla2008numerical} for benchmarking purposes. The same approach can in principle be used for more complicated bare model systems as described in Sec.~\ref{sec:complex}. However, we recognize that useful applications of UML may require integration with \textit{ab initio} methods to treat realistic systems \cite{foulkes2001quantum,bartlett2007coupled,sugiyama1986auxiliary,motta2018ab,zhu2024towards,calvo2024theoretical}. This is beyond the scope of the current work, but we discuss possible ways to do this as an outlook in Appendix~\ref{app:abinitio}.

The UML framework that we have introduced is based on unsupervised ML techniques, with the goal to optimize parameters of a variational effective Hamiltonian, to best describe the low-temperature physics of a more complex bare model quantum impurity system. In a well-defined sense, the optimized effective Hamiltonian is the ``best fit'' description at a given complexity level. We propose a loss function to quantify the distinguishability between bare and effective models, based on a classical distribution of quantum Feynman diagrams within a hybridization expansion of the partition function. The loss function can be minimized using GD by computing certain physical observables. In the standard case we discuss, the resulting optimization problem is convex, resulting in a highly efficient learning process. The optimized effective model is the one in which local static observables match in bare and effective models.

Care must be taken however, since thermodynamic observables are not invariant under RG and so bare an effective models with different RG flows may have different static expectation values.  Thus, the naive matching of observables does not imply that low-energy scales, such as the Kondo temperature, are correctly recovered. We confirm this directly, already at the simplest possible example of the Anderson-Kondo model mapping. To mitigate this we identify and construct the best observables to match, based on an analysis of their RG relevance. We show that this can be done while maintaining convexity of the optimization, and that low-energy scales can be recovered. 
The accuracy of our effective models can be systematically improved by increasing the complexity of the effective impurity. \rev{Although this clearly offers no advantage at the level of the AIM, UML produces dramatically simplified effective models when applied to more complex systems, consisting of 3 or more interacting active orbitals. 
The accuracy and efficiency of UML may be further improved by formulating bare and effective models in the `Natural Orbital' (NO) basis \cite{yang2017unveiling}. Since entanglement between NOs and the rest of the fermionic bath is small by construction, local observables to be matched in UML should flow weakly, which we have showed is a desirable property when formulating effective models. }

An important aspect of the UML framework is that it can be performed at relatively high temperatures without compromising precision. This opens the door to applying the algorithm to vastly more complex systems than could be treated by brute-force NRG calculations alone. For example, a one-shot CT-QMC calculation of observables for the bare model could be used as a reference to optimize a simpler effective model treated cheaply within NRG, which can then be solved down to $T=0$. Integration with \textit{ab-initio} methods may permit the realistic study of complex strongly correlated systems that are currently out of reach.

%########################
%########################

\begin{acknowledgments}
The authors would like to thank Sudeshna Sen, Alexander Munoz, Lucas K.~Wagner and Denjoe O'Connor for the insightful discussion. We acknowledge funding from the Irish Research Council Laureate Awards 2017/2018 through grant IRCLA/2017/169 (JBR and AKM), from Science Foundation Ireland through grant 21/RP-2TF/10019 (AKM) and from the Helmholtz Initiative and Networking Fund, grant no.~VH-NG-1711 (JBR).
\end{acknowledgments}

\appendix

%########################
%########################

\section{Hybridization expansion and convexity of the loss function}
\label{app:convexity}

For a general Hamiltonian, that can be bi-partitioned $ \hat{H} =  \hat{H}_0 +  \hat{H}_1$, one can compute the partition function as an expansion in powers of $ \hat{H}_1$ by \cite{Haldane_1978},
\begin{align}
	\label{eq:expansion1}
	\mathcal{Z} = \sum^\infty_{n=0} \frac{(-1)^n}{n!}T_\tau&\int^\beta_0\dd\tau_1...\int^\beta_{0}\dd\tau_n \nonumber \\
	&\times\tr\left[e^{-\beta  \hat{H}_0} \hat{H}_1(\tau_n)... \hat{H}_1(\tau_1)\right] \; ,
\end{align}
where $\tau$ is the imaginary time. This expansion is discussed extensively in the context of continuous time quantum Monte Carlo (CT-QMC) \cite{PhysRevB.72.035122,RevModPhys.83.349,PhysRevB.74.155107}. In the case of hybridization expansion CT-QMC \cite{PhysRevB.74.155107} $ \hat{H}_{1} =  \hat{H}^{\rm{hyb}}$ describes the hybridization between a non-interacting bath $\hat{H}^{\rm{bath}}$ and an interacting quantum impurity $ \hat{H}^{\rm{imp}}$. For the following discussion, we therefore consider Hamiltonians of the form,
\begin{subequations}
	\begin{align}
		\label{eq:impham}
		\hat{H}_0 &=  \hat{H}^{\rm{bath}} +  \hat{H}^{\rm{imp}} \\
		\hat{H}_1 &=  \hat{H}^{\rm{hyb}} = \sum_{k}\sum_{\sigma} V^{\sigma}_{k} \hat{d}^{\dagger}_{\sigma} \hat{c}_{k \sigma} + \text{H.c.} \; ,
	\end{align}
\end{subequations}
where we assume that the hybridization tensor is diagonal in the spin quantum number $\sigma$ \cite{PhysRevB.74.155107}. For this type of Hamiltonian \ceqn{eq:expansion1} becomes,
\begin{align}
	\label{eq:rawexp}
	\mathcal{Z} = &\sum^\infty_{n=0} \int^{\beta}_{0}\dd\tau_1...\int^{\beta}_{\tau_{n-1}}\dd\tau_n\int^{\beta}_0\dd\tau'_1...\int^{\beta}_{\tau'_{n-1}}\dd\tau'_n  \nonumber \\ &\sum_{\substack{a_1 ... a_n\\ a'_1 ... a'_n}} \sum_{\substack{k_1 ... k_n\\ k'_1 ... k'_n}} V^{a_1}_{k_1}V^{a'_1 *}_{k'_1}...V^{a_n}_{k_n}V^{a'_n *}_{k'_n}\\
	&\tr\left[ T_\tau e^{-\beta \hat{H}^{\rm{bath}}} c^\dagger_{k'_n a'_n}(\tau'_n)\hat{c}_{k_n a_n}(\tau_n) ... \hat{c}^\dagger_{k'_1 a'_1}(\tau'_1)\hat{c}_{k_1 a_1}(\tau_1)\right]  \nonumber\\
	\times &\tr\left[ T_\tau e^{-\beta \hat{H}^{\rm{imp}}} \hat{d}_{a'_k}(\tau'_n)\hat{d}^\dagger_{a_n}(\tau_n) ... \hat{d}_{a'_1}(\tau'_1)\hat{d}^\dagger_{a_1}(\tau_1)\right] \nonumber\; .
\end{align}
Eq.~\ref{eq:rawexp} can be interpreted as sum over all possible diagrams obtained by allowing electrons to hop between the impurity and the bath. Since the bath is non-interacting it can be integrated out and using Wick's theorem the \textit{antiperiodic hybridization function} can be obtained, 
\begin{align}
	&\det_{i j}\left[V^{\sigma_i}_{k_i}V^{\sigma'_j *}_{k'_j}\tr \left( T_{\tau} e^{-\beta  \hat{H}^{\rm{bath}}} \hat{c}^{\dagger}_{k_i \sigma_i}(\tau_i)\hat{c}_{k'_j \sigma'_j}(\tau'_j)\right)\right] \nonumber \\
	&=\mathcal{Z}^{\rm{bath}}\det_{i j}\left[V^{\sigma_i}_{k_i}V^{\sigma'_j *}_{k'_j}\langle T_{\tau}\hat{c}^{\dagger}_{k_i \sigma_i}(\tau_i)\hat{c}_{k'_j \sigma'_j}(\tau'_j)\rangle^{\rm{bath}} \right] \nonumber \\
	&= \mathcal{Z}^{\rm{bath}} \det( \Delta^{(x)}) \; ,
\end{align}
where $x = (n,\lbrace k_i, k'_i,a_i,a'_i,\sigma_{i}, \sigma'_{i},\tau_{i}, \tau'_{i}\rbrace^{n}_{i=1})$ denotes an impurity diagram in terms of the sequence of impurity operators \cite{Haldane_1978} such that,
\begin{equation}
	\label{eq:mulint}
	\int^{\beta}_{0}\dd\tau_1...\int^{\beta}_{\tau_{n-1}}\dd\tau_k\sum_{a_1 ... a_n}\sum_{k_1 ... k_n}\mapsto \sum_x \;.
\end{equation}
Following the approach of Ref.~\cite{haule2007quantum}, we bring the impurity operators into the eigenbasis $\lbrace (E_{\alpha},\ket{\alpha}) \rbrace$ of $ \hat{H}^{\rm{imp}}$ and thus, $e^{-\tau \hat{H}^{\rm{imp}}}$ can be trivially evaluated, with
\begin{equation}
	\label{eq:deig}
	\begin{gathered}
		d^{\dagger}_{a}(\tau) = e^{- \tau_i \hat{H}^{\rm{imp}}}d^{\dagger}_{a}e^{\tau_i \hat{H}^{\rm{imp}}} \\
		= \sum_{\alpha,\alpha'}e^{\tau(E_{\alpha'} - E_\alpha)}\ket{\alpha}\!\!\bra{\alpha'}\bra{\alpha} d^{(\dagger)}_{a} \ket{\alpha'}
	\end{gathered}
\end{equation}
and instead of occupation diagrams we obtain diagrams involving impurity eigenstates $\lbrace \alpha \rbrace_x \equiv \lbrace \alpha_1 ... \alpha_k, \alpha'_1 ... \alpha'_k \rbrace_x$. Thus, \ceqn{eq:expansion1} can be rewritten in terms of the weights of the sum of the eigenstate diagrams
\begin{align}
	\label{eq:expansion3}
	\mathcal{Z}/\mathcal{Z}^{\rm{bath}} &= \sum_{x} \sum_{\lbrace\alpha\rbrace} 
	 \Lambda_{\lbrace \alpha \rbrace_x}\det(\Delta_x)
	e^{-\langle \hat{H}^{\rm{imp}}\rangle_{\lbrace \alpha \rbrace_x}} \;,
\end{align}
with $\Lambda_{\lbrace \alpha \rbrace_x}$ denoting the contribution from impurity operators in the impurity eigenbasis 
\begin{equation}
	\Lambda_{\lbrace \alpha \rbrace_x} =  \prod^{n}_{i=1} \bra{\alpha_{i}}d_{a'_{i}}\ket{\alpha'_{i}}\bra{\alpha'_{i}}d^{\dagger}_{a_{i}}\ket{\alpha_{i+1}} 
\end{equation}
and $\langle \hat{H}^{\rm{imp}}\rangle_{\lbrace \alpha \rbrace_x}$ the average impurity energy over a diagram
\begin{gather}
\langle \hat{H}^{\rm{imp}}\rangle_{\lbrace \alpha \rbrace_x} = \\ \sum^{n}_{i=1}\left[ E_{\alpha_{i}}(\tau'_{i} - \tau_{i-1}) + E_{\alpha'_{i}}(\tau_{i}-\tau'_{i}) \right] + E_{\alpha_n}(\beta - \tau_n ) \nonumber \; .
\end{gather}
Having reformulated the expansion of the impurity partition \ceqn{eq:rawexp} as \ceqn{eq:expansion3} we now look to prove the \textcolor{new_text}{convexity} of KLD loss function \ceqn{eq:P_KLD}.
 
%%%%%%%%
% new convexity proof
%%%%%%%%
To show the convexity of \ceqn{eq:P_KLD} it does suffice to show that the free energy $\log(\mathcal{Z})$ is convex in the variational parameters $\pmb{\theta}$. The core assumption of our calculation is that the effective impurity Hamiltonian is constructed as follows:
\begin{equation}
\label{eq:convex_assumptions}
H^{\rm imp} = \sum_i \theta_i \hat{h}_i,\qquad [\hat{h}_i,\hat{h}_j] = 0\;,
\end{equation}
which is the case for the effective models in the main text. 
For the following proof it is convenient to define a \textit{weight} $w$ for eigenstate diagrams to reformulate \ceqn{eq:expansion3},
\begin{gather}
	\mathcal{Z} =\mathcal{Z}^{\rm bath} \sum_{x} \sum_{\lbrace\alpha\rbrace_x} 
	w(\lbrace \alpha \rbrace_x) \;,\\
	w(\lbrace \alpha \rbrace_x) = e^{-\langle \hat{H}^{\rm{imp}}\rangle_{\lbrace \alpha \rbrace_x}}
	\Lambda_{\lbrace \alpha \rbrace_x}\det(\Delta_x) \;,
\end{gather}
where we also assume that $w(\lbrace \alpha \rbrace_x) > 0$, such that $w$ acts as a probability distribution upon normalization. 
To show that $\log(\mathcal{Z})$ is convex we need to show its Hessian is positive semi-definite,
\begin{gather}
\partial_{\theta_i} \partial_{\theta_j} \log(\mathcal{Z}) \succcurlyeq 0 \;.
\end{gather}
In the following, we use the shorthand notation $\partial_{i}$ for $\partial_{\theta_i}$ for concision. 
It is straightforward to compute the first order derivative using the fact that the trace is invariant under circular shifts,
\begin{equation*}
	\begin{aligned}
		\partial_i \ln(\mathcal{Z}) &= -\frac{1}{\mathcal{Z}}\int^\beta_0 d\tau \operatorname{tr}\left[T_\tau e^{-\int^\tau_0 d\tau' \hat{H}} \:\hat{h}_i(\tau) \: e^{-\int^\beta_\tau d\tau' \hat{H} }\right] \\
		&= -\frac{1}{\mathcal{Z}}\int^\beta_0 d\tau \operatorname{tr}\left[e^{-\int^\tau_0 d\tau' \hat{H} - \int^\beta_\tau d\tau' \hat{H} } ~ \hat{h}_i(\tau)\right] \\
		&= -\frac{1}{\mathcal{Z}}\int^\beta_0 d\tau \sum_\alpha e^{-\beta E_\alpha} \bra{\alpha}e^{-\tau E_\alpha} \hat{h}_i \: e^{\tau E_\alpha}\ket{\alpha} \\
		&= -\beta\operatorname{tr}[\hat{\rho} ~ \hat{h}_i] 
		= -\beta\langle \hat{h}_i \rangle \; .
	\end{aligned}
\end{equation*}
However we can also carry out the derivative directly on the diagrammatic expansion of the partition function,
\begin{align}
	\partial_i \log(\mathcal{Z}) &= \frac{1}{\mathcal{Z}}\partial_i\bigg[\mathcal{Z}^{\rm bath} \sum_{x} \sum_{\lbrace\alpha\rbrace_x}e^{-\langle \hat{H}^{\rm{imp}}\rangle_{\lbrace \alpha \rbrace_x}}
	\Lambda_{\lbrace \alpha \rbrace_x}\det(\Delta_x)\bigg] \nonumber\\
	&= \frac{1}{\mathcal{Z}}\bigg[\mathcal{Z}^{\rm bath} \sum_{x} \sum_{\lbrace\alpha\rbrace_x}\partial_i w(\lbrace \alpha \rbrace_x) \bigg]  \;.
\end{align}
It holds that a given eigenbasis $\lbrace (E_{\alpha},\ket{\alpha}) \rbrace$ of $ \hat{H}^{\rm{imp}}$ depends in the following way on the parameters $\pmb{\theta}$, 
\begin{align}
\partial_i E_\alpha &= \bra{\alpha}\hat{h}_i\ket{\alpha} \;, \\
\partial_i \ket{\alpha} &= \sum_{\alpha \neq \beta }\frac{\bra{\beta}\hat{h}_i\ket{\alpha}}{E_\alpha - E_\beta}\ket{\beta} \;.
\end{align}
With the assumptions \ceqn{eq:convex_assumptions} in place all operators in the Hamiltonian are mutually commuting, implying that they share a common set of eigenvectors. This property allows us to trivially evaluate these derivatives,
\begin{gather}
\partial_i E_\alpha = \epsilon^\alpha_i\\
\partial_i \ket{\alpha}  = 0\;,
\end{gather}
with $\hat{h}_i \ket{\alpha} =  \epsilon^\alpha_i\ket{\alpha}$. With these relationships it is straightforward to show that,
\begin{align}
\partial_j \Lambda_{\lbrace \alpha \rbrace_x} &= 0 \\
\partial_j \langle \hat{H}^{\rm{imp}}\rangle_{\lbrace \alpha \rbrace_x} &= \langle \hat{h}_j\rangle_{\lbrace \alpha \rbrace_x}\;.
\end{align}
This allows the calculation of the derivative of $w$, which amounts to,
\begin{gather}
-\partial_i w(\lbrace \alpha \rbrace_x) = \langle \hat{h}_i\rangle_{\lbrace \alpha \rbrace_x} w(\lbrace \alpha \rbrace_x) \;.
\end{gather}
This shows that $w$ is a generating functional of moments of $\langle \hat{h}_i\rangle_{\lbrace \alpha \rbrace_x}$. We therefore write,
\begin{align}
-\partial_i \log(\mathcal{Z})
&= -\frac{1}{\mathcal{Z}}\bigg[\mathcal{Z}^{\rm bath} \sum_{x} \sum_{\lbrace\alpha\rbrace_x}\partial_i w(\lbrace \alpha \rbrace_x) \bigg] \\
&= \mathbb{E}_w\big[ \langle \hat{h}_i\rangle_{\lbrace \alpha \rbrace_x}\big] \;.
\end{align}
Accordingly the second order derivative becomes,
\begin{gather}
\partial_j\partial_i \log(\mathcal{Z})
= \mathbb{E}_w\big[ \langle \hat{h}_i\rangle_{\lbrace \alpha \rbrace_x} \langle \hat{h}_j\rangle_{\lbrace \alpha \rbrace_x}\big] \\ -\mathbb{E}_w\big[ \langle \hat{h}_i\rangle_{\lbrace \alpha \rbrace_x}\big]\mathbb{E}_w\big[ \langle \hat{h}_j\rangle_{\lbrace \alpha \rbrace_x}\big] \;,
\end{gather}
which can be identified as the covariance. Since the covariance is a positive-semidefinite matrix, we have,
\begin{gather}
\partial_j\partial_i \log(\mathcal{Z}) = \operatorname{cov}_w\big[\langle \hat{h}_i\rangle_{\lbrace \alpha \rbrace_x} ,\, \langle \hat{h}_j\rangle_{\lbrace \alpha \rbrace_x}\big]  \succcurlyeq 0 \; ,
\end{gather}
making the Hessian of $\log(\mathcal{Z})$ positive semi-definite and the same applies to the Hessian of the KLD loss function \ceqn{eq:P_KLD} under the aforementioned assumptions \ceqn{eq:convex_assumptions}. 

%############

\section{Extended impurity basis}
\label{app:double}
To construct the projectors of the extended MC Hamiltonian, we require the basis $\ket{Q,S,T,S_z;m}$ on the extended Fock space Eq.~\ref{eq:Fockext}.  The first step in the construction is to determine the charge $\hat{Q}$, total spin $\TS^2$ and $\hat{S}_z$ eigenstates $\ket{Q,S,S_z;m}$. For the occupation basis we use the labelling convention: 
\begin{equation}
	\ket{\rm{orbital~2}~(L),\rm{orbital~1}~(R),\rm{local~ moment}~(M)} \; .
\end{equation}
Thus, the Hilbert space is spanned by the states:
\begin{widetext}
\begin{center}
	\begin{tabular}{ |c|c|c| } 
		$\ket{-2,1/2,S_z;m}$ & $\ket{-1,1,S_z;m}$ & $\ket{-1,0,S_z;m}$ \\
		\hline
		$\ket{0,0,\Upa}$ & $\ket{\upa,0,\Upa}$&~$1/\sqrt{2}(\ket{\upa,0,\Doa}-\ket{\doa,0,\Upa})$~ \\
		$\ket{0,0,\Doa}$ & $\ket{\doa,0,\Doa}$ & $1/\sqrt{2}(\ket{0,\upa,\Doa}-\ket{0,\doa,\Upa})$ \\
		& ~$1/\sqrt{2}(\ket{\upa,0,\Doa}+\ket{\doa,0,\Upa})$~ & \\
		& $\ket{0,\upa,\Upa}$ & \\
		& $\ket{0,\doa,\Doa}$ & \\
		& $1/\sqrt{2}(\ket{0,\upa,\Doa}+\ket{0,\doa,\Upa}) $ &
	\end{tabular}
\end{center}

\begin{center}
	\begin{tabular}{ |c|c|c|} 
		$\ket{0,3/2,S_z;m}$ & $\ket{0,1/2,S_z;m}$ & $\ket{0,1/2,S_z;m}$\\
		\hline
		$\ket{\upa,\upa,\Upa}$ & $\sqrt{2/3}\ket{\upa,\upa,\Doa} - \sqrt{1/6}(\ket{\doa,\upa,\Upa}+\ket{\upa,\doa,\Upa})$ & $\ket{\ud,0,\Upa}$\\
		~$1/\sqrt{3}(\ket{\doa,\upa,\Doa}+\ket{\doa,\doa,\Upa}+\ket{\upa,\doa,\Doa})$~ &  
		$~-\sqrt{2/3}\ket{\doa,\doa,\Upa} + \sqrt{1/6}(\ket{\doa,\upa,\Doa}+\ket{\upa,\doa,\Doa})$& $\ket{\ud,0,\Doa}$~ \\
		$1/\sqrt{3}(\ket{\doa,\upa,\Upa}+\ket{\upa,\upa,\Doa}+\ket{\upa,\doa,\Upa})$ & & $\ket{0,\ud,\Upa}$ \\
		$\ket{\doa,\doa,\Doa}$ &
		& $\ket{0,\ud,\Doa}$ \\
		& & ~$\sqrt{1/2}(\ket{\doa,\upa,\Upa}- \ket{\upa,\doa,\Upa})$~ \\
		& & $\sqrt{1/2}(\ket{\doa,\upa,\Doa}- \ket{\upa,\doa,\Doa})$
	\end{tabular}
\end{center}
\end{widetext}
The states $\ket{2,1/2,S_z;m},\ket{1,1,S_z;m}$ and $\ket{1,0,S_z;m}$ can be obtained straightforwardly by replacing all $0$ entries with $\ud$. It is important to note that when using only $Q$ and $S$ quantum numbers, the $Q = -1,S =1$ subspace is two-fold degenerate and the $Q = 0, S=1/2$ subspace is threefold degenerate. Therefore, using only the projectors onto the $Q$ and $S$ quantum number subspaces limits the expressibility of our effective model. Full tunability of the model requires that we can address each basis state individually. The next step is thus to lift these degeneracies. We may do this by diagonalizing the hopping operator in the degenerate subspaces,
\begin{equation}
	\hat{T} = \sum_\sigma \hat{c}^\dagger_{L\sigma}\hat{c}_{R\sigma} + \hat{c}^\dagger_{R\sigma}\hat{c}_{L\sigma} \; .
\end{equation}
In the $\ket{QS}= \ket{\pm1,1}$ and $\ket{\pm1,0}$ sub-spaces we can find a common eigenbasis $\ket{Q,S,T,S_z;m}$ of $\hat{T}$ and $\TS^2$ that allows to completely lift the multiplet degeneracy, e.g.~
\begin{widetext}
\begin{center}
	\centering
	\begin{tabular}{ |c|c| } 
		$\ket{-1,1,+1,S_z;m}$ & $\ket{-1,1,-1,S_z;m}$  \\
		\hline
		$1\sqrt{2}(\ket{\upa,0,\Upa}+\ket{0,\upa,\Upa})$ & $1\sqrt{2}(\ket{\upa,0,\Upa}-\ket{0,\upa,\Upa})$ \\
		~$1/\sqrt{4}(\ket{0,\upa,\Doa}+\ket{0,\doa,\Upa}+\ket{\upa,0,\Doa}+\ket{\doa,0,\Upa})$~ & ~$1/\sqrt{4}(\ket{0,\upa,\Doa}+\ket{0,\doa,\Upa}-\ket{\upa,0,\Doa}-\ket{\doa,0,\Upa})$~ \\
		$1\sqrt{2}(\ket{\doa,0,\Doa}+\ket{0,\doa,\Doa})$&$1\sqrt{2}(\ket{\doa,0,\Doa}-\ket{0,\doa,\Doa})$
	\end{tabular}
\end{center}
\end{widetext}
%Here we have used that
%\begin{align*}
%	\hat{T}\ket{0,\doa} = - \ket{\doa,0} \; .
%\end{align*}
%Thus, the symmetric combination of $\ket{-1(1),1(0),S_z;m}$ states is associated to the negative $T=-1$ eigenvalue of $\hat{T}$ and the asymmetric combination to the positive eigenvalue $T=1$ of $\hat{T}$.

In the $Q = 0$ subspace one can show that the hopping has no effect on the quadruplet and the doublet. However the $Q = 0, S =1/2$ subspace behaves non-trivially under the hopping:
\begin{flalign*}
	\hat{T} \ket{\ud,0,\Upa} &= \ket{\doa, \upa, \Upa} - \ket{\upa,\doa,\Upa} \\
	\hat{T} \ket{0,\ud,\Upa} &= \ket{\doa, \upa, \Upa} - \ket{\upa,\doa,\Upa} \\
	\hat{T}( \ket{\doa, \upa, \Upa} - \ket{\upa,\doa,\Upa} )/\sqrt{2} &= \sqrt{2}( \ket{\ud,0,\Upa} +  \ket{0,\ud,\Upa})
\end{flalign*}
We can combine these states such that they become eigenstates of $\hat{T}$:
\begin{widetext}
\begin{adjustwidth}{-1.5cm}{-1.5cm}
	\begin{center}
		\centerline{%begin centerline
		\begin{tabular}{ |c|c|c|} 
			$\ket{0,1/2,0,S_z;m}$ & $\ket{0,1/2,+2,S_z;m}$&$\ket{0,1/2,-2,S_z;m}$\\
			\hline
			$\sqrt{\frac{2}{3}}\ket{\upa,\upa,\Doa} - \sqrt{\frac{1}{6}}(\ket{\doa,\upa,\Upa}+\ket{\upa,\doa,\Upa})$ & 
			$\frac{1}{\sqrt{4}}(\ket{\doa, \upa, \Upa} - \ket{\upa,\doa,\Upa}  + \ket{\ud,0,\Upa} +  \ket{0,\ud,\Upa})$&
			$\frac{1}{\sqrt{4}}(\ket{\doa, \upa, \Upa} - \ket{\upa,\doa,\Upa}  - \ket{\ud,0,\Upa} -  \ket{0,\ud,\Upa})$\\
			$-\sqrt{\frac{2}{3}}\ket{\doa,\doa,\Upa} + \sqrt{\frac{1}{6}}(\ket{\doa,\upa,\Doa}+\ket{\upa,\doa,\Doa})$& 
			$\frac{1}{\sqrt{4}}(\ket{\doa, \upa, \Doa} - \ket{\upa,\doa,\Doa}  + \ket{\ud,0,\Doa} +  \ket{0,\ud,\Doa})$&
			$\frac{1}{\sqrt{4}}(\ket{\doa, \upa, \Doa} - \ket{\upa,\doa,\Doa}  - \ket{\ud,0,\Doa} -  \ket{0,\ud,\Doa})$ \\
			$\frac{1}{\sqrt{2}}( \ket{\ud,0,\Upa} -  \ket{0,\ud,\Upa})$  & & \\
			$\frac{1}{\sqrt{2}}( \ket{\ud,0,\Doa} -  \ket{0,\ud,\Doa})$  & &
		\end{tabular}
	}%ending center line
	\end{center}
\end{adjustwidth}
\end{widetext}
The \textit{fourfold} degenerate multiplet is now almost completely lifted. Only the $\ket{0,1/2,0,S_z;m}$ subspace is still twofold degenerate. This last degeneracy can be lifted by introducing a further operator,
\begin{equation*}
	\hat{W} = \hat{n}^L_{\upa} \hat{n}^L_{\doa} + \hat{n}^R_{\upa} \hat{n}^R_{\doa} \; ,
\end{equation*}
which has the eigenvales $W \in \lbrace 0,1 \rbrace$ in the $\ket{0,1/2,0,S_z;m}$ subspace. Instead of introducing a new label for only this subspace, the eigenvalues for $\hat{T}$ and $\hat{W}$ are added together in the $T$ label.  This gives the states:
\begin{widetext}
	\begin{adjustwidth}{-1.5cm}{-1.5cm}
		\begin{center}
			\begin{tabular}{ |c|c|} 
				$\ket{0,1/2,1,S_z;m}$ & $\ket{0,1/2,0,S_z;m}$\\
				\hline
				$\sqrt{\frac{2}{3}}\ket{\upa,\upa,\Doa} - \sqrt{\frac{1}{6}}(\ket{\doa,\upa,\Upa}+\ket{\upa,\doa,\Upa})$ & $\frac{1}{\sqrt{2}}( \ket{\ud,0,\Upa} -  \ket{0,\ud,\Upa})$ \\
				$-\sqrt{\frac{2}{3}}\ket{\doa,\doa,\Upa} + \sqrt{\frac{1}{6}}(\ket{\doa,\upa,\Doa}+\ket{\upa,\doa,\Doa})$& $\frac{1}{\sqrt{2}}( \ket{\ud,0,\Doa} -  \ket{0,\ud,\Doa})$ 
			\end{tabular}
		\end{center}
	\end{adjustwidth}
\end{widetext}
The dimension of each subspace spanned by these labels has thereby been reduced to $M=1$, as desired.

%#########

\section{Details of NRG calculations}
\label{app:NRG}

All calculations in this work were carried out using Wilson's NRG method \cite{wilson1975the,RevModPhys.80.395}
using the full density matrix approach \cite{PhysRevLett.95.196801,PhysRevLett.99.076402}, which allows the precise calculation of the static observables discussed here. The impurity entropy, which is used to estimate $T_K$ via $S_{\rm imp}(T_K)=\tfrac{1}{2}$, was computed using standard thermodynamic NRG \cite{wilson1975the}. For the simulation of bare and effective models, the number of kept states is $M_K=2000$ and we use a Wilson chain discretization parameter of $\Lambda = 2.5$ at a Wilson chain length of $N = 40$. Total charge and spin projection quantum numbers are exploited.

%%%%%%%%%%%%%%%%%%%
%############\textbf{}

\section{Integration with\\\textit{ab initio} methods}\label{app:abinitio}
At the heart of the UML method is the computation of local observables in bare and effective models. The effective models are simple enough that this can be done accurately at any temperature using e.g.~NRG.  However, evaluating objects such as $\langle \hat{P}_{QS} \rangle$ in the bare model within an \textit{ab initio} description is in general challenging. 

For methods that operate in second quantization, we present here a straightforward way to construct the bare projectors $\hat{P}_{QS}$ that project onto the quantum number states $\ket{Q,S}$. First we introduce the auxiliary operators,
\begin{align}
	\hat{X}_Q &= \hat{Q} - Q \\
	\hat{Y}_S &= \hat{\mathbf{S}}^{2} - S(S+1) \; ,
\end{align}
where $\hat{\mathbf{S}}^{2}$ is the total spin operator of the extended impurity and $\hat{Q}$ the total charge operator of the extended impurity.
The operator $\hat{X}_Q$ eliminates all contributions with charge $Q$ to a state $\ket{\psi}$ and similarly $\hat{Y}_S$ eliminates all contributions with total spin $S$.
Using these operators we can now write the projector onto the multiplet subspace of $Q$ and $S$ as
\begin{align}
	\label{eq:pqs-rdm}
	\hat{P}_{QS} = \frac{1}{N_{QS}}\prod_{Q' \neq Q} \hat{X}_{Q'} \times \prod_{S' \neq S} \hat{Y}_{S'} \; ,
\end{align}
where 
\begin{align}
	N_{QS} = \prod_{Q' \neq Q} (Q - Q') \prod_{S' \neq S} (S(S+1) - S'(S'+1)) \; ,
\end{align}
is a normalization constant. One can similarly construct the admissibility operator $\hat{\Omega}_{\rm ad}$. However, these operators are manifestly of high order: Eq.~\ref{eq:pqs-rdm} is an $N$-body operator, where $N$ is the number of particles in the extended bare impurity Hilbert space.

Within an \textit{ab initio} framework, expectation values of such $N$-body operators are typically intractable, because they are usually expressed in terms of reduced density matrices (RDM) \cite{mazziotti2012two}.  For a given basis of molecular orbitals $\lbrace \phi_i \rbrace^M_{i=1}$ one can express the spinless single particle RDM (1-RDM) as
\begin{equation}
	\rho_1(r';r) = \sum^M_{i,j} ~^1 D^i_{j} \phi_i ( r' )\phi_j (r)\;,
\end{equation}
where $~^1 D^i_{j}= d_i \times d_j = \langle \psi | \hat{c}^\dagger_i \hat{c}_j | \psi \rangle$ is the 1-RDM which  encodes a single particle wave-function as,
\begin{equation}
	\psi(r) = \sum^M_i d_i \phi_i(r)  \;  .
\end{equation}
The 1-RDM, is not a many body object, although it is related to the 2-RDM $~^2 D$ as \cite{mazziotti2012two}
\begin{equation}
	~^2 D^{pq}_{st} = 2~^1D^s_p \wedge \!~^1D^q_t + ~^2\Delta^{pq}_{st}\;
\end{equation}
where  $~^2\Delta$ is a cumulant that encodes the relations between multiple determinants. The tensor $D^{pq}_{st} = \langle \psi | \hat{c}^\dagger_p \hat{c}^\dagger_q \hat{c}_s \hat{c}_t | \psi \rangle $ encodes two-body correlations. The construction of the $N$-RDM is therefore possible for an arbitrary number for particles $N$, but it becomes prohibitively expensive rather quickly.

Instead of attempting to compute \ceqn{eq:pqs-rdm} exactly within this framework, we instead sketch a method for reducing the projectors $\hat{P}_{QS}$ to a simplified but approximate form that might be compatible with \textit{ab initio} methods using 1- or 2-body RDMs.

We wish to find the best variational approximate representation to \ceqn{eq:pqs-rdm} using at most 2-body operators. We can do this using the FSP, \ceqn{eq:forbenius}. We therefore construct a full basis of operators for the bare Hilbert space,
\begin{equation}
	\label{eq:basisx}
	\hat{A} \in \Xi_n \in \lbrace \lbrace\hat{c}^\dagger_{i}\hat{c}_j\rbrace,~\lbrace\hat{c}^\dagger_{i}\hat{c}^\dagger_{j}\hat{c}_k\hat{c}_l\rbrace,~\lbrace\hat{c}^\dagger_{i}\hat{c}^\dagger_{j}\hat{c}^\dagger_{k}\hat{c}^\dagger_{l}\hat{c}_m\hat{c}_n\hat{c}_o\hat{c}_p\rbrace,\dots\rbrace \; .
\end{equation}
Now retaining only the 1- and 2-body terms, we have a basis $\{\hat{h}_i\}$ for the approximate representation of $\hat{P}_{QS}$.  For a complete basis $\Xi_n $, the decomposition $\hat{P}_{QS}=\sum_i \theta_i\hat{h}_i$ is unique and can be achieved like a regular spectral decomposition of a vector using the FSP, viz:
\begin{equation}
	\theta_i = \frac{\langle \hat{P}_{QS}, \hat{h}_i\rangle}{\norm{\hat{h}_i}}\;, \qquad\qquad\hat{h}_i \in \Xi_n \; .
\end{equation}
However, we note that the basis \ceqn{eq:basisx} is generally over-complete and therefore any representation of $\hat{P}_{QS}$ according to \ceqn{eq:forbenius} is not necessarily unique. Constructing a complete operator basis on the bare Hilbert space can be challenging. On the other hand, the variational approach can find optimal representations of any operator even for an over-complete basis, and the value of the optimized loss function can serve as an error estimation for the decomposition. We argue that this is a controlled approach to obtain 2-body approximations of the UML operators $\hat{P}_{QS}$. We leave development and implementation of these methods for future work.

%########################
%########################

\bibliography{refs}

%########################
%########################

\end{document}